\documentclass[a4paper,fleqn]{cas-sc}

\usepackage[authoryear,longnamesfirst]{natbib}
\usepackage{graphicx} 
\usepackage[linesnumbered,ruled,norelsize,algo2e]{algorithm2e}

 \usepackage{algorithm}
 \usepackage{tikz}
\usepackage{xspace}
\usepackage{float}
\usepackage[dvipsnames]{xcolor}
\usepackage[many]{tcolorbox}
\usetikzlibrary{calc}
\tcbuselibrary{skins}
\usepackage{booktabs,tabularx}

\usepackage{hyperref}  

\newcommand{\algCE}{\textsf{SMT-PLF}\xspace}
\newcommand{\algOP}{\textsf{Opt-PLF}\xspace}
\newcommand{\algEL}{\textsf{Quadratic}\xspace}

\allowdisplaybreaks

\newcommand{\R}{{\mathbb{R}}}

\newcommand{\f}{\mathcal{F}}
\newcommand{\g}{\mathcal{G}}
\newcommand{\ii}{\mathcal{I}}
\newcommand{\V}{\mathbb{V}}
\newcommand{\D}{\mathbb{D}}
\newcommand{\N}{{\mathbb{N}}}
\newcommand{\Q}{\mathbb{N}_q}

\newcommand{\TO}{\mathcal{T}}
\newcommand{\argmin}{\mathrm{arg}\min} 
\newcommand{\co}{\mathrm{Co}}
\newcommand{\ra}{\rightarrow}
\newcommand{\Ra}{\Rightarrow}
\newcommand{\Let}{:=}

\newcommand{\dI}{\Delta \ii}
\newcommand{\dr}{\Delta R}
\newcommand{\Ld}{\hat{L}}
\newcommand{\Ad}{\hat{A}}
\newcommand{\Nu}{\mathcal{V}}
\newcommand{\overbar}[1]{\mkern 1.5mu\overline{\mkern-1.5mu\text#1\mkern-1.5mu}\mkern 1.5mu}
\newproof{pf}{Proof}
\newcounter{theorem}
\newcounter{definition}
\newcounter{lemma}
\newcounter{claim}
\newcounter{problem}
\newcounter{proposition}
\newcounter{corollary}
\newcounter{construction}
\newcounter{example}
\newcounter{xca}
\newcounter{comments}
\newcounter{remark}
\newcounter{assumption}

\newtheorem{theorem}[theorem]{Theorem}
\newtheorem{lemma}[lemma]{Lemma}
\newtheorem{problem}[problem]{Problem}

\newtheorem{definition}[definition]{Definition}

\newtheorem{remark}[remark]{Remark}

\numberwithin{equation}{section}

\newtcolorbox{resp1}[1][]{%
	enhanced jigsaw,%
	colback=gray!5!white,%
	colframe=gray!80!black,%
	size=small,%
	boxrule=1pt,%
	halign title=flush center,%
	coltitle=black,%
	breakable,%
	drop shadow=black!50!white,%
	attach boxed title to top left={xshift=1cm,yshift=-\tcboxedtitleheight/2,yshifttext=-\tcboxedtitleheight/2},%
	minipage boxed title=3cm,%
	boxed title style={%
		colback=white,%
		size=fbox,%
		boxrule=1pt,%
		boxsep=2pt,%
		underlay={%
			\coordinate (dotA) at ($(interior.west) + (-0.5pt,0)$);
			\coordinate (dotB) at ($(interior.east) + (0.5pt,0)$);
			\begin{scope}[gray!80!black]
				\fill (dotA) circle (2pt);
				\fill (dotB) circle (2pt);
			\end{scope}
		}%
	},%
	#1%
}

\def\tsc#1{\csdef{#1}{\textsc{\lowercase{#1}}\xspace}}
\tsc{WGM}
\tsc{QE}

\begin{document}
\let\WriteBookmarks\relax
\def\floatpagepagefraction{1}
\def\textpagefraction{.001}

\shorttitle{Policy Synthesis for Interval MDPs via Polyhedral Lyapunov Functions}    

\shortauthors{N. Monir \& S. Soudjani}  

\title [mode = title]{Policy Synthesis for Interval MDPs via Polyhedral Lyapunov Functions}  

\tnotemark[1] 


%

\author[1]{Negar Monir}

\cormark[1]

\fnmark[1]

\ead{s.seyedmonir2@newcastle.ac.uk}



\affiliation[1]{organization={School of Computing, Newcastle University},
            city={Newcastle upon Tyne},
            country={United Kingdom}}

\author[2,3]{Sadegh Soudjani}

\fnmark[2]

\ead{sadegh@mpi-sws.org}



\affiliation[2]{organization={Max Planck Institute for Software Systems},
            country={Germany}}

\affiliation[3]{organization={University of Birmingham},
            city={Birmingham},
            country={United Kingdom}}

\cortext[1]{Corresponding author}

\fntext[1]{The research of N. Monir is supported by the EPSRC EP/W524700/1 and Newcastle University Global Scholarship.}

\fntext[2]{The research of S.~Soudjani is supported by the following grants: EIC 101070802 and ERC 101089047.}


\begin{abstract}
Decision-making under uncertainty is central to many safety-critical applications, where decisions must be guided by probabilistic modeling formalisms.
This paper introduces a novel approach to policy synthesis in multi-objective interval Markov decision processes using polyhedral Lyapunov functions. Unlike previous Lyapunov-based methods that mainly rely on quadratic functions, our method utilizes polyhedral functions to enhance accuracy in managing uncertainties within value iteration of dynamic programming. We reformulate the value iteration algorithm as a switched affine system with interval uncertainties and apply control-theoretic stability principles to synthesize policies that guide the system toward a desired target set. By constructing an invariant set of attraction, we ensure that the synthesized policies provide convergence guarantees while minimizing the impact of transition uncertainty in the underlying model. Our methodology removes the need for computationally intensive Pareto curve computations by directly determining a policy that brings objectives within a specified range of their target values. We validate our approach through numerical case studies, including a recycling robot and an electric vehicle battery, demonstrating its effectiveness in achieving policy synthesis under uncertainty. 
\end{abstract}




\maketitle

\section{Introduction}\label{sec:Introduction}
\noindent\textbf{Motivation.}  
Decision-making under uncertainty poses challenges in various domains such as autonomous systems, robotics, and healthcare. Markov decision processes (MDPs) provide a modeling framework for such problems, allowing agents to choose actions based on probabilistic state transitions to optimize an objective function \citep{bellman1957markovian}. However, many applications require addressing multiple competing objectives. For example, self-driving cars must balance safety and efficiency, energy grids weigh cost against reliability, and medical diagnostics often involve a trade-off between accuracy and interpretability. This complexity has led to the development of Multi-objective MDP (MOMDP) frameworks, which aim to create policies that balance these competing criteria \citep{10.1007/11672142_26, etessami2008multi}. An added complexity arises when transition probabilities are uncertain due to factors like sensor noise, environmental changes, or incomplete knowledge. Interval MDPs (IMDPs) address this by modeling transition probabilities as intervals instead of fixed values \citep{GIVAN200071, HADDAD2018111}. Numerous techniques have been developed to synthesize optimal or near-optimal policies for IMDPs, often providing guarantees for worst-case performance \citep{nilim2005robust, WU2008945, Wolff2012, delimpaltadakis2023interval}. To further improve decision-making, Multi-objective IMDPs (MOIMDPs) have been studied \citep{hahn2017multi, hahn2019interval}, necessitating specialized algorithms that manage uncertain transitions and multi-objective queries while ensuring policy robustness. 

Dynamic programming effectively solves optimization problems in MDPs. Key techniques include value iteration (VI) and policy iteration. VI iteratively updates state values until reaching convergence, yielding an optimal value function that helps identifying the best policy for each state. This process streamlines computations, making VI a powerful tool for decision-making in probabilistic systems \citep{bertsekas2011dynamic, DELGADO2016192}. Due to the challenges of large MDPs, researchers are exploring approximate dynamic programming and reinforcement learning to improve scalability \citep{bertsekas2011approximate, sutton2018reinforcement, lavaei2023compositional}. Furthermore, studies have also looked into their convergence properties and theoretical limitations \citep{NIPS1996_e0040614, tsitsiklis2002convergence, ha2021generalized}.

Recent research has concentrated on extending policy synthesis based on dynamic programming to multi-objective settings \citep{10.1007/11672142_26, etessami2008multi, hahn2017multi, hahn2019interval} and incorporating interval-based uncertainty in MDPs \citep{HADDAD2018111, mathiesen2024intervalmdp}. However, previous studies that tackled both multi-objective optimization and interval-based transitions have primarily relied on computing Pareto curves to identify policies that balance different objectives \citep{hahn2017multi, hahn2019interval, scheftelowitsch2017multi}.
As mentioned in \citep{hahn2019interval}, it is generally impossible to derive an exact representation of the Pareto curve in polynomial time, and an $\epsilon$-approximation is necessary to compute it. This motivates the research of our paper: we propose a novel Lyapunov-based VI algorithm by representing the VI as a discrete-time switched affine system (dt-SAS) and utilizing polyhedral Lyapunov functions (PLF) to analyze convergence and synthesize policies.

\smallskip
\noindent\textbf{Related Works.}
Using dynamical systems theory to analyze the convergence of iterative algorithms has been studied in the literature. This includes gradient descent for first-order optimization \citep{194420, brockett1991dynamical, helmke2012optimization}, the steepest descent algorithm in computer vision \citep{bloch1990steepest, bloch1990kahler}, and learning in deep neural networks \citep{yeung2019learning, rajendra2020modeling,  xie2022new}.
A key application of this methodology in policy synthesis is the study by \cite{doi:10.1080/00207179.2021.2005260}, which has used Lyapunov theory to study the convergence of VI in MDPs by formulating dynamic programming as a dt-SAS and utilizing an ellipsoidal invariant set of attraction (ISoA) to stabilize the dt-SAS. However, it is limited to single-objective MDPs and does not account for interval uncertainty, limiting its applicability to IMDPs and multi-objective settings. A recent study by \cite{tipaldi2025data} has recast VI for MDPs as a dt-SAS, incorporating model uncertainty by treating transition and reward matrices as unknown, but it is limited to single-objectives and has not considered interval uncertainties.

To stabilize dt-SAS, several Lyapunov-based methods have been employed, particularly through defining an ISoA that confines system trajectories to a region of attraction. A common approach involves using an ellipsoidal ISoA from a quadratic Lyapunov function \citep{Deaecto2016}. The study by \cite{albea2020robust} introduces a robust ellipsoidal ISoA for policy synthesis, though this design requires relaxations that may enlarge the stability region and reduce accuracy. A scenario-based extension studied by \cite{monir2024switch} and adapted from \citet{Deaecto2016} for uncertain settings tightens the ISoA but increases computational costs. These trade-offs necessitate the exploration of less conservative stabilization certificates for dt-SAS.

Quadratic Lyapunov functions are commonly used for stability analysis, but can result in conservative stability regions due to their symmetry and fixed curvature, limiting policy synthesis. Recent studies suggest PLFs as a more flexible and accurate alternative for representing the polyhedral ISoA \citep{sun2011stability, blanchini2008set, ahmadi2016lower}. PLFs are piecewise affine and can be tailored to the system's geometry, making them less conservative in defining stability regions. This advantage is particularly relevant in switched and hybrid systems, where dynamics vary across different modes, necessitating non-uniform stability guarantees \citep{polanski2000absolute, berger2022data}. Additionally, PLFs facilitate counterexample-guided approaches \citep{ahmed2020automated, berger2022learning, berger2023counterexample}.

\begin{figure}
    \centering
    \includegraphics[width=.6\linewidth]{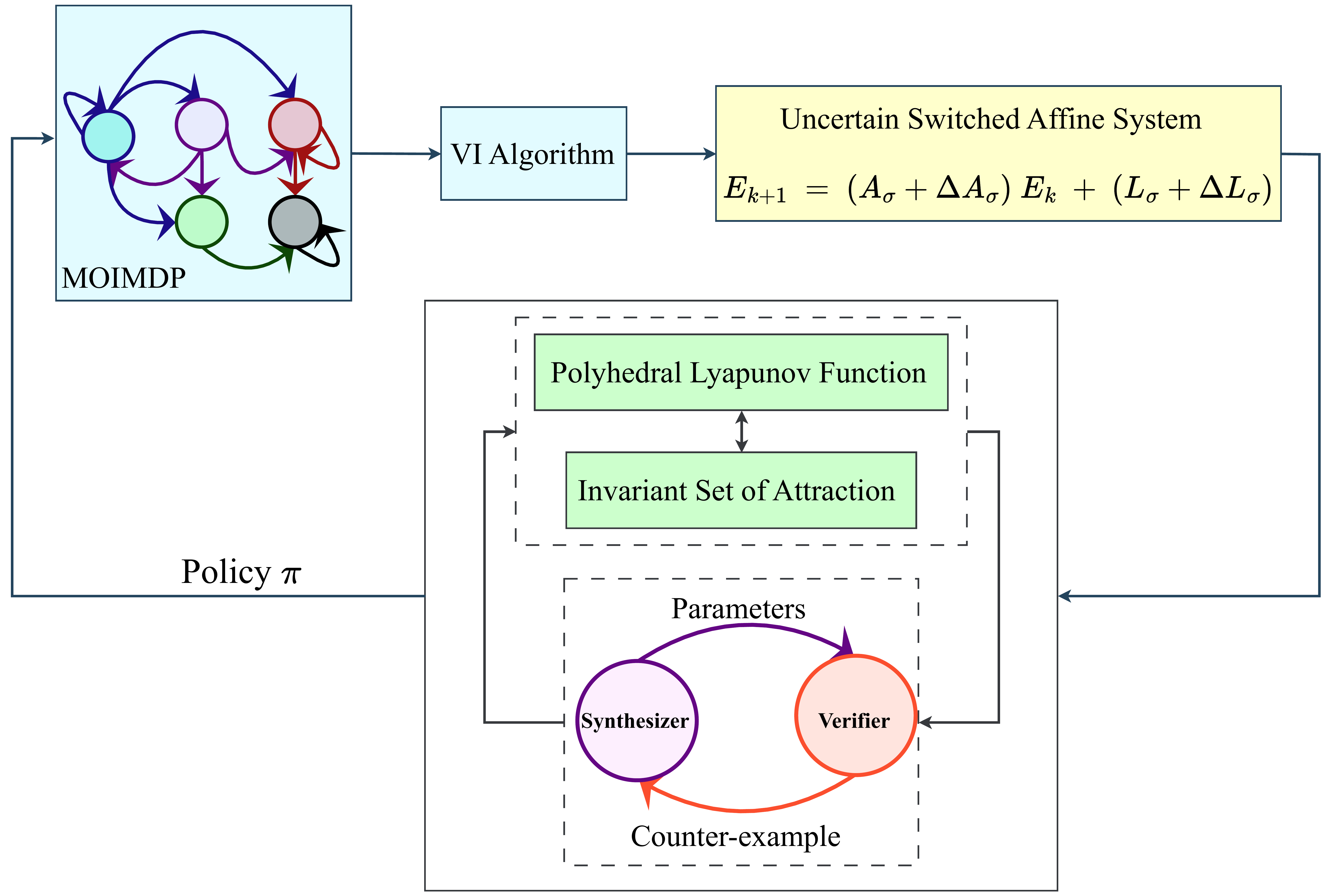}
    \caption{High-level representation of the proposed approach.}
    \label{fig:HL Rep}
\end{figure}

\smallskip
\noindent\textbf{Contributions.}
This paper introduces a novel Lyapunov-based approach for policy synthesis in MOIMDPs by utilizing PLFs. This method enhances the accuracy of defining stability regions and synthesizing policies. We model MOIMDPs as a dt-SAS with interval uncertainty, which enables us to apply Lyapunov stability theory to ensure that the system converges toward the desired target values. In contrast to previous Lyapunov-based VI methods that depended on quadratic Lyapunov functions—resulting in conservative stability regions—our approach employs PLFs. This enables us to create a tighter ISoA, thereby improving the precision of policy synthesis while maintaining robustness under uncertainty. Additionally, we utilize a counterexample-guided algorithm to iteratively refine the parameters of the PLFs and their corresponding ISoA. This ensures a systematic and adaptive process for stability verification. A high-level representation of our proposed algorithm is illustrated in Figure~\ref{fig:HL Rep}, which outlines the key steps involved in policy synthesis and stability analysis.

A preliminary version of the approach of this paper has been presented in the conference paper \citep{monir2024lyapunov}, which utilizes quadratic Lyapunov functions and non-convex optimizations, thus not aligned with the piecewise affine structure of the state equations. The current manuscript extends the conference version in the following directions:
    (a) We provide a policy-synthesis framework based on PLFs that produces tighter and more accurately shaped invariant sets. While ellipsoidal certificates derived from quadratic Lyapunov functions create regions of attraction that are ellipsoidal in shape, this often results in over-smoothing and over-approximating the actual geometry of switched or affine systems. In contrast, polyhedral Lyapunov functions define regions with adjustable facets that can align with switching boundaries and dominant directions. This added geometric flexibility eliminates the restriction of an ellipsoidal shape, reduces conservatism, and leads to more precise policies in uncertain environments.
    (b)
    Since PLFs support counterexample-guided computation, we propose two counterexample-guided algorithms to design PLFs. This approach refines stability conditions based on the constraints violations, leading to more accurate stability verification.

\smallskip
\noindent\textbf{Outline.}
The structure of this paper is as follows. Section~\ref{sec:pre}
presents the preliminaries and the problem statement. In Section~\ref{sec: Model SAS}, we model the MOIMDP as a dt-SAS with uncertainties. Section~\ref{sec policy synth} presents two counterexample-guided algorithms to design a polyhedral ISoA for the uncertain dt-SAS together with a robust VI algorithm to design a policy for the MOIMDP. In Section~\ref{sec: case studies}, we present case studies that demonstrate the effectiveness of our approach and compare the results with the baseline approach. Finally, Section~\ref{sec: conclusion} provides a conclusion to the paper with suggestions for future research.

\section{Preliminaries and Problem Statement}
\label{sec:pre}
In this section, we present key definitions, preliminaries, and the problem statement.

\smallskip
\noindent\textbf{Notations.}
$\N$,  $\R$, $\R_{>0}$, $\R_{\geq 0}$, $\R^{n}$, and $\R^{n \times m}$ denote, respectively, the sets of natural numbers {including zero}, real numbers, positive real numbers, non-negative real numbers, the n-dimensional Euclidean space, and the set of $n \times m$ real matrices. $\N_m$ denotes the set $\{1,2,\dots, m\}$ and $0_{n\times m}$ indicates the $n\times m$ matrix with zero elements. The cardinality of any set $A$ is denoted by $|A|$. 
For any $M \in \R^{n \times n}$, $\det(M)$ represents the determinant of $M$. For any matrices $A=A^{\top}$, $B$, and $C=C^{\top}$ of appropriate dimensions, we abbreviate $$\left[\begin{array}{ll}A & B \\ * & C\end{array}\right]:=\left[\begin{array}{cc}A & B \\ B^{\top} & C\end{array}\right].$$ $\co(A, B)$ is the convex hull generated by matrices $A$ and $B$. The interior of a set is represented by 
$\operatorname{int}(\cdot)$. The concatenation of two matrices $A$ and $B$ of appropriate dimensions in a row is denoted by $[A,B]$ and in a column by $[A;B]$. Finally, the unitary simplex of dimension $m\in\N$ is defined as $\Lambda_m:=\left\{\left[\lambda_{1}, \ldots, \lambda_m\right] \mid \lambda_{i}\ge 0,\sum_{i=1}^m \lambda_{i}=1\right\}$. 

\subsection{Multi-objective Interval Markov Decision Processes}
We study the dynamic programming framework for optimizing multiple objectives on interval Markov decision processes (MOIMDP).
\begin{definition}
An \emph{MOIMDP} is a tuple \( \Sigma = (X, x_0, U, \underbar{P}, \overbar{P},\) \(\underbar{R}, \overbar{R}) \), comprising a finite set of states \( X \), an initial state \( x_0 \in X \), and a finite set of actions \( U \).
Lower and upper bounds on the transition probabilities between the states are given by
$\underbar{P},\overbar{P}: X \times U \times X \rightarrow [0,1]$. Lower and upper bounds on a collection of $q$ reward functions are given by $\underbar{R}, \overbar{R}:X \times U \rightarrow \mathbb R^q$, where $\underbar{R}:=[\underbar{r}_1;\underbar{r}_2,\ldots;\underbar{r}_q]$ and $\overbar{R}:=[\overbar{r}_1;\overbar{r}_2,\ldots;\overbar{r}_q]$. 
\label{def: moimdp}
\end{definition}
For any state-action pair $(x,u)\in X\times U$, the consecutive state $x'\in X$ is selected according to some probability distribution $x'\sim \ii(\cdot|x,u)$, where $\underbar{P}(x'|x,u)\leq \ii(x'|x,u)\leq \overbar{P}(x'|x,u)$ for all $x,u,x'$. {For any $(x,u)\in X\times U$, the rewards $r_m(x,u)$, $m\in\Q$, will belong to the intervals $[\underbar{r}_m,\overbar{r}_m]$.} 
A \emph{Markov} policy $\boldsymbol{\pi} = (\pi_0,\pi_1,\pi_2,\ldots)$ is a sequence of functions $\pi_t: X \ra U$ that map states into actions at any time $t\in\N$. We denote by $\Pi$ the set of all such policies. Note that for finite $X$ and $U$, the set of possible functions $\pi:X\ra U$ is also finite, denoted by $\bar{\Pi}=\{\pi^1,\pi^2,\ldots,\pi^M\}$ with $M =|U|^{|X|}$.
The policy $\boldsymbol{\pi}$ is called \emph{stationary} if there is $\pi:X \ra U$ such that $\boldsymbol{\pi}=(\pi,\pi,\pi,\ldots)$. In this case, with abuse of notation, we simply use $\pi$ to denote the stationary policy. 

Any policy $\boldsymbol{\pi}$ together with the choice of probability distributions $\ii(x'|x,u)$ at each time step induces a probability measure on the sequence of states $x(0),x(1),x(2),\ldots$. 
The performance of a policy $\boldsymbol{\pi}$ is assessed with respect to its induced expected total discounted reward, which also depends on the choice of transition probabilities $\ii(x'|x,u)$ from the interval $[\underbar{P}(x'|x,u),\overbar{P}(x'|x,u)]$ and rewards from the interval $[\underbar{R}(x,u),\overbar{R}(x,u)]$. Therefore, it is essential to study the best- and worst-case performance with respect to the uncertainties in the transition probabilities and in the rewards.

For a given policy $\boldsymbol{\pi}$ and discount factors $0 < \gamma_i < 1$, $i \in \N_q$, define the best- and worst-case performance of the expected total discounted reward as a function of an arbitrary initial state as 
\begin{align*}
w_{\boldsymbol{\pi},\textsf{opt}}^m(x_0) :=
      \textsf{opt}_{\boldsymbol{\mu}} \mathbb E^{\boldsymbol{\pi},\boldsymbol{\mu}}\!\left[\sum_{t=0}^{\infty} \!\gamma_m^tr_m(x(t),\pi_t(x(t)))|x(0)\!=\!x_0\right],
\end{align*}
where $\textsf{opt}\in\{\max,\min\}$ is taken with respect to an adversarial policy $\boldsymbol{\mu}$ that chooses the transition probabilities $\ii$ from the interval $[\underbar{P},\overbar{P}]$, and the rewards $r_m$ from the interval $[\underbar{r}_m,\overbar{r}_m]$. The expectation operator $\mathbb E^{\boldsymbol{\pi},\boldsymbol{\mu}}$ is with respect to the probability distribution induced on the sequence of states under the policies $(\boldsymbol{\pi},\boldsymbol{\mu})$. We denote these value functions by $w_{\pi,\textsf{opt}}^m(\cdot)$ when the policy $\boldsymbol{\pi}$ is stationary, which satisfy the following Bellman equation \citep{bertsekas2011dynamic}
\begin{align*}
    w_{\pi,\textsf{opt}}^m(x)= \textsf{opt}_{\ii,r_m} \Big[r_m(x,\pi(x))
    + \sum_{x^{\prime} \in X }\gamma_m\ii(x'|x,\pi(x)){w}_{\pi,\textsf{opt}}^m(x^{\prime})\Big],
\end{align*}
for all $x\in X$ and $m \in \Q $.

Assuming that $X$ has $n$ states and for some fixed order on the states, the value function can be represented as a vector, which with abuse of notation, is indicated also by $w_{\pi,\textsf{opt}}^m \in \R^n$. The Bellman operator can also be defined in vector form as $\TO_\pi:\R^n\ra\R^n$ with
\begin{equation*}
\TO_{\pi} w = \textsf{opt}_{\ii,R_m}\left[R_m(\pi)+\gamma_m \ii(\pi)w\right], \quad\forall w\in \R^n,
\end{equation*}
where $R_m(\pi)\in \R^n$ is a vector containing the reward intervals for $r_m(x,\pi(x))$, and
$\ii(\pi)$ is the matrix containing transition probability intervals under $\pi$; each element of the vector $R_m$ and the matrix $\ii$ can be written as intervals $R_{mi} \in [\underbar{R}_{mi}, \overbar{R}_{mi}]$ and  $\ii_{ij} \in [\underbar{P}_{ij}, \overbar{P}_{ij}]$, respectively. The VI to calculate $w_{\pi,\textsf{opt}}^m$ is
\begin{equation}
    w_{k+1}^m=\TO_{\pi} w_{k}^m, \quad \forall k\in\N \text{ with } w_{0}^m = 0, 
    \label{VI Op moimdp}
\end{equation}
where $w_{\pi,\textsf{opt}}^m=\lim_{k\ra \infty}w_{k}^m$.
The VI~\eqref{VI Op moimdp} needs to be solved twice with $\textsf{opt}\in\{\max,\min\}$ to obtain the best- and worst-case performance with respect to the uncertainties in the transition probabilities and in the rewards. 

\begin{remark}
    [Multi-objective MDPs]
If $\underbar{P}=\overbar{P}=P$ and $\underbar{R} =  \overbar{R}$ in Definition~\ref{def: moimdp}, we obtain an MOMDP with transition probabilities $P$ and rewards $R$, denoted by \( \Sigma = (X, x_0, U, P, R) \). By substituting $P$ and $R$ in equations of MOIMDP, dropping the optimization $\textsf{opt}$ with respect to uncertainties, and considering value functions $v_{\pi}^m$, we get the affine Bellman operator for MOMDPs as
\begin{equation*}
\TO_{\pi} v = R_m(\pi)+\gamma_m P(\pi)v, \quad\forall v\in \R^n.
\end{equation*}
The corresponding VI algorithm to calculate $v_{\pi}^m$ is
\begin{equation}
v_{k+1}^m=\TO_{\pi} v_{k}^m, \quad \forall k\in\N \text{ with } v_{0}^m = 0,
\label{update rule convergence momdp}
\end{equation}
where $v_{\pi}^m=\lim_{k\ra \infty}v_{k}^m$.
\end{remark}
\subsection{Problem Statement}
Given an MOIMDP \( \Sigma = (X, x_0, U, \underbar{P}, \overbar{P}, \underbar{R}, \overbar{R}) \), we aim to design a policy $\boldsymbol{\pi}$ such that the value function in~\eqref{VI Op moimdp} converges to a neighborhood of a target value  $W_{tar}$.
\begin{resp1}
\begin{problem}
\label{moimdp prblm}
    Given an MOIMDP $\Sigma$ and target value $W_{tar}:=[w_{tar}^1;w_{tar}^2;\dots;w_{tar}^q]$,
    find a set $\g$ and a policy $\boldsymbol{\pi}$ such that $W_{tar}\in \g$ and $W_{\boldsymbol{\pi},\textsf{opt}} = [w_{\boldsymbol{\pi},\textsf{opt}}^1;w_{\boldsymbol{\pi},\textsf{opt}}^2;\dots;w_{\boldsymbol{\pi},\textsf{opt}}^q]\in\g$.
\end{problem}
\end{resp1}
The above problem is trivially feasible for any $W_{tar}$ and any stationary policy $\pi$ by taking a sufficiently large $\g$. This paper aims to find a set $\g$ that is as tight as possible using a novel Lyapunov-based VI algorithm as described in the next section. 
\section{VI Algorithm as dt-SAS with Uncertainties}
\label{sec: Model SAS}

To address Problem~\ref{moimdp prblm}, we reformulate the VI algorithm in~\eqref{VI Op moimdp} for the MOIMDP $\Sigma = (X, x_0, U, \underbar{P}, \overbar{P}, \underbar{R},\overbar{R})$ as a stability problem for a dt-SAS with uncertainties (dt-USAS) that models the underlying error dynamics.
Then, we design a polyhedral ISoA using counterexample guided computation in order to synthesize a robust switching policy that drives the value function towards the desired target values.

Let $W_{k}=[w_{k}^1;w_{k}^2;\ldots;w_{k}^q]$ be the augmented value function for all q objectives. Thus, the solution of the VI algorithm in~\eqref{VI Op moimdp} is included in the set of solutions of 
\begin{equation*}
    W_{k+1}= A_{\pi_k} W_{k} + B_{\pi_k} , 
\end{equation*} 
with interval matrices
\begin{align}
    A_{\pi}=\mathrm{diag}(\gamma_1 \ii(\pi),\ldots,\gamma_q \ii(\pi)), \text{ and }
    B_{\pi}= \big[\begin{matrix}R_1(\pi);R_2(\pi); \ldots; R_q(\pi)\end{matrix}\big], \label{uncertain matrices A B}
\end{align}
for all $\pi\in\bar{\Pi}$. In addition, we define the error w.r.t. target values $w_{tar}^m$ as $e_{k}^m:=w_k^m-w_{tar}^m$ for all objectives $m \in \Q$, which has the dynamics
\begin{equation*}
e_{k+1}^m = \gamma_m \ii(\pi_k)e_k^m + l_{\pi_k}^m, 
\end{equation*}
where $l_{\pi}^m := R_m(\pi)-(I-\gamma_m \ii(\pi))w_{tar}^m$.
Similar to the augmented value function $W_k$,
the augmented error $E_{k}=[ e_{k}^1; e_{k}^2; \ldots; e_{k}^q]$ is described by the switched affine system subject to uncertainty
\begin{equation}
    E_{k+1} = A_{\pi_k}E_{k} + L_{\pi_k}, \label{Augmented E moimdp}
\end{equation}
where $L_{\pi_k}= [ l_{\pi_k}^1; l_{\pi_k}^2; \ldots;l_{\pi_k}^q] = B_{\pi_k} - (I-A_{\pi_k})W_{tar}$. Note that both matrices $A_{\pi_k}$ and $L_{\pi_k}$ have uncertainties, which can be expressed as a convex hull
\begin{equation}
    [A_\pi,L_\pi] \in \co ([A_\pi^\kappa,L_\pi^\kappa])_{\kappa \in \D},\quad  \forall \pi \in\bar{\Pi}, \label{politopic uncertainity moimdp}
\end{equation}
generated by a finite $\D$ number of matrices $A_\pi^j$ and $L_\pi^j$.

\subsection{Constructing a Feasible Set of Values \label{feasible solution momdp subsecc}}
In this subsection, we consider finding a feasible set of target values in Problem~\ref{moimdp prblm} by analyzing the asymptotic stability of the evaluation of a stationary policy. We determine the steady-state values that can be reached by the value functions under a nominal model.

We rewrite $\ii$ that has elements $\ii_{ij}\in[\underbar{P}_{ij},\overbar{P}_{ij}]$ as $\ii = \hat{\ii} + \dI$ with $\hat{\ii}$ being a nominal transition probability matrix for the IMDP. One option would be to choose the elements $\hat{\ii}_{ij}$ as close as possible to $\frac{1}{2}(\underbar{P}_{ij} + \overbar{P}_{ij})$ and $|\dI_{ij} (\pi)| \leq \Delta_{ij}$
such that $ [\hat{\ii}_{ij}-\Delta_{ij},\hat{\ii}_{ij}+\Delta_{ij}]\supset[\underbar{P}_{ij},\overbar{P}_{ij}]$.
Similarly, we rewrite $R$ that has elements $R_{ij}\in[\underbar{R}_{ij},\overbar{R}_{ij}]$ as $R = \hat{R} + \dr$ with $\hat{R}$ being a nominal reward for the IMDP.
One option would be to choose $\hat{R}_{ij} = \frac{1}{2}(\underbar{R}_{ij} + \overbar{R}_{ij})$.
Note that $(X, x_0, U, , \hat{\ii}, \hat{R})$ is an MOMDP as the nominal model.
Also, define the nominal matrices
\begin{align}
    \Ad_\pi := \mathrm{diag}(\gamma_1 \hat{\ii}(\pi),\ldots,\gamma_q \hat{\ii}(\pi)),\text{ and } \hat{B}_\pi := \left[\hat{R}_1(\pi);\hat{R}_2(\pi);\ldots;\hat{R}_q(\pi)\right].\label{nominal matrices}
\end{align}
The VI for the nominal model is
\begin{equation}
\label{eq:nominal_VI}
   \hat W_{k+1}= \hat A_{\pi_k} \hat W_{k} + \hat B_{\pi_k}, 
\end{equation}
which admits the steady state values $\hat W_{\pi,ss}$
that satisfy
\begin{equation}
    \hat W_{\pi,ss} = \hat B_\pi + \hat A_\pi \hat W_{\pi,ss}. \label{augmented steady update rule convergence momdp}
\end{equation}
Note that $\hat A_\pi$ is always invertible since $\gamma_m \in (0,1)$ for all $m \in \Q$ and the spectral radius of $\hat{\ii}(\pi)$ is equal to 1 for any $\pi$. Here, we can rewrite~\eqref{augmented steady update rule convergence momdp} as follows
\begin{equation*}
    \hat W_{\pi,ss} = (I-\hat A_\pi)^{-1}\hat B_\pi.
\end{equation*}
Using $\hat W_{\pi,ss}$ as target values $W_{tar}$, we get that $\Ld_\pi := \hat{B}_{\pi} - (I-\Ad_{\pi})W_{tar} = 0$. Therefore, the error dynamics of the nominal model become 
\begin{equation}
\label{eq:nominal_error}
   \hat E_{k+1} =  \hat A_\pi \hat E_{k},
\end{equation}
which is a classical linear time-invariant discrete-time model. The nominal error will converge to zero regardless of the initial error (i.e., the model is globally asymptotically stable having all the eigenvalues of $\hat A_\pi$ inside the unit circle in the complex plane).
\begin{lemma}
For any stationary policy $\pi$, any $\gamma_m\in(0,1)$ with $m \in \Q$, and any initial value $\hat W_{0}$,  $\hat W_k$ in~\eqref{eq:nominal_VI} asymptotically converges to the corresponding steady-state values $\hat W_{\pi,ss} = (I-\hat A_\pi)^{-1}\hat B_\pi$.
\label{convergence lemma momdp}
\end{lemma}
We study Problem~\ref{moimdp prblm} for guiding the value functions $W_k$ towards a given target value $W_{tar} =[ 
    w_{tar}^1;
    w_{tar}^2;
    \ldots;
    w_{tar}^q
]$.
Consider the following feasible set of generalized equilibrium points that are specified with respect to the nominal matrices:
\begin{equation}
    \f \Let \{W \in \R^{qn} : W=(I-{\Ad}_{\lambda})^{-1}\hat{B}_{\lambda},
    \lambda \in \Lambda_{M}\}, \label{feasable set imdp}
\end{equation}
where $\Lambda_{M}$ is the unitary simplex with dimension $M$ and
\begin{equation}
    {\Ad}_{\lambda}:=\sum_{\pi \in \bar{\Pi}}\lambda_{\pi}{\Ad}_{\pi}, \text{ and }  \hat{B}_{\lambda}=\sum_{\pi \in\bar{\Pi}}\lambda_{\pi}\hat{B}_{\pi}. \label{Aland Blanda imdp}
\end{equation}
These generalized equilibrium points are computed as the equilibrium points resulting from the convex combinations of the dynamics in~\eqref{eq:nominal_VI} associated with the policies $\pi \in\bar{\Pi}$. 
For any $W_{tar}\in\f$, the error dynamics are the same as in~\eqref{Augmented E moimdp}.
\subsection{Stability Analysis of the dt-USAS}
In this subsection, we analyze the stability of the dt-USAS described in~\eqref{Augmented E moimdp} using Lyapunov-based theories. We aim to create an ISoA derived from the PLF, which guarantees that the error dynamics converge to a designated area surrounding the origin. The goal is to construct a switching law \(\pi(E)\) such that, ideally, $E_k \rightarrow 0$ as $k \rightarrow \infty$ for all initial conditions. However, in the presence of uncertainty in the system, achieving this condition exactly is generally not feasible. Instead, we design \(\pi(E)\) to switch between elements of \(\bar{\Pi}\) to guide the trajectories of the system toward the ISoA that includes the origin. To formally analyze stability and construct this set, we give the definition of the ISoA.
\begin{definition}[ISoA]
\label{invariant set mdp def}
    A set $\Omega \subset \R^{qn}$ is an \emph{Invariant Set of Attraction (ISoA)} of the system~\eqref{Augmented E moimdp} governed by the switching policy $\pi_k$, if there exists a Lyapunov function $\V:\R^{qn}\rightarrow \R_{\ge 0}$ such that the following conditions are satisfied simultaneously: 
\begin{enumerate}[(C\arabic*)]
    \item\label{C1} $0 \in \Omega$, 
    \item\label{C2} If ${E}_k \notin \Omega$, then $\Delta \V({E}_k) := \V({E}_{k+1})-\V({E}_k) < 0$,
    \item\label{C3} If ${E}_k \in \Omega$, then ${E}_{k+1} \in \Omega$. 
\end{enumerate}
\end{definition}

If such a Lyapunov function exists, it is guaranteed that under the switching policy, the error will always converge to the set $\Omega$ for any initial error $E_0$ and will stay inside $\Omega$ afterwards. The system is then called practically stable \citep{deaecto2016practical}.

Based on Definition~\ref{invariant set mdp def}, we employ the candidate PLF
\begin{equation}
\label{eq. poly lya}
    \V(E) := 
    \max_{c \in \Nu} \left({c}^\top E-d_c\right)
 \end{equation}
 for some finite set of vectors $\mathcal V$ and real values $d_c$ allowing us to shift the center of the level sets. 
According to condition~(C2), the following minimum-type state feedback switching control law is selected:
\begin{equation}
\label{eq. switch law}
    \pi(E) = \argmin_{\pi \in \bar{\Pi}} \V(A_\pi E + L_\pi).
\end{equation}
We also take the level set of $\V$
\begin{equation}
    \Omega \Let \{E \in \R^{qn} : \V({E}) \leq \rho\}, \label{relax invariant set momdp}
\end{equation}
as the candidate polyhedral ISoA, for some $\rho$ to be designed.

\begin{resp1}
\begin{problem}
\label{poly prblm}
    Given the error dynamics of the dt-USAS in~\eqref{Augmented E moimdp}, design the parameters of the PLF in~\eqref{eq. poly lya} and $\rho>0$ such that $\Omega$ in~\eqref{relax invariant set momdp} becomes an ISoA for the dt-USAS under the switching law~\eqref{eq. switch law} according to Definition~\ref{invariant set mdp def}. 
\end{problem}
\end{resp1}

To solve this problem, we provide algorithmic solutions in the next section. 

\section{Policy Synthesis for MOIMDPs} \label{sec policy synth}
This section addresses both Problems~\ref{moimdp prblm}-\ref{poly prblm} by proposing two algorithms for designing polyhedral ISoA using counterexample guided inductive synthesis (CEGIS) \citep{abate2018counterexample}, and by providing a robust VI algorithm for synthesizing a policy. 

At a high level, the CEGIS approach operates by alternating between proposing a candidate certificate that meets the current set of requirements and querying a verifier. The verifier either returns a counterexample or confirms the validity of the candidate certificate; this process continues until no counterexamples are found. The first algorithm is described in Subsection~\ref{sec cegis} utilizing Satisfiability Modulo Theories (SMT) solvers \citep{barrett2018satisfiability}. The second algorithm described in Subsection~\ref{sec opt learn} is optimization-based, where certificates and counterexamples are determined by solving optimization problems instead of using SMT solvers. 
Furthermore, a robust VI algorithm is proposed in Subsection~\ref{sec: robust vi alg}. This algorithm utilizes either of the computed PLFs and their polyhedral ISoA to solve Problem~\ref{moimdp prblm}.
\subsection{Synthesizing Polyhedral ISoA via CEGIS and SMT}
\label{sec cegis}
In this section, we set $\rho=1$ for the ISoA in~\eqref{relax invariant set momdp} without loss of generality, and establish the conditions on the parameters of the PLF to give a solution for Problems~\ref{poly prblm}.

\begin{theorem}
\label{thm cegis}
The function $\V(E)$ in \eqref{eq. poly lya},
the switching policy $\pi$ in \eqref{eq. switch law}, and
the ISoA $\Omega$ in \eqref{relax invariant set momdp} with $\rho=1$ give a solution for Problem~\ref{poly prblm} if the parameters of $\V(E)$ satisfy the following statement:
    \begin{align}
    &\Psi: \forall E,\,\exists \pi,\,\{(\forall c_i \in \Nu, \, d_{c_i} \ge -1)\,\wedge\,(\exists c_i \in \Nu,{c_i}^\top E-d_{c_i}\ge 0)\,\wedge\,([\forall c_i \in \Nu, (c_i^\top E-d_{c_i}) \le 1] \,\,\vee \nonumber \\ 
    &[\forall c_j \in \Nu, \forall \kappa \in \D, \exists c_\ell \in \Nu, (c_j^\top (A_\pi^\kappa E + L_\pi^\kappa) - d_{c_j} < c_\ell^\top E- d_{c_\ell})]) \, \wedge \, ([\exists c_i \in \Nu,\, (c_i^\top E -d_{c_i}) > 1] \,\vee\, \nonumber \\
    &[\forall \kappa \in \D,\forall c_j \in \Nu, c_j^\top (A_\pi^\kappa E + L_\pi^\kappa)-d_{c_j} \leq 1])\}. \label{eq pr PSI}
\end{align}
\end{theorem}
The proof of the theorem is relegated to the appendix. Note that the negation of the condition $\Psi$ can be written explicitly as
\begin{align}
    &\neg \Psi: \exists E,\,\forall \pi,\, \{(\exists c_i \in \Nu, \, d_{c_i} < -1)\,\vee\,( \forall c_i \in \Nu, {c_i}^\top E-d_{c_i} < 0) \,\vee\, ([\exists c_i \in \Nu, (c_i^\top E-d_{c_i}) > 1] \,\wedge \nonumber \\ 
    &[\exists c_j \in \Nu, \exists \kappa \in \D, \forall c_\ell \in \Nu, (c_j^\top (A_\pi^\kappa E + L_\pi^\kappa) - d_{c_j} \ge c_\ell^\top E- d_{c_\ell})])\,\vee\,([\forall c_i \in \Nu,\, (c_i^\top E -d_{c_i}) \le 1] \,\wedge\, \nonumber\\
    &[\exists \kappa \in \D,\exists c_j \in \Nu, c_j^\top (A_\pi^\kappa E + L_\pi^\kappa)-d_{c_j} > 1])\}. \label{eq pr neg Psi}
\end{align}
Although Theorem~\ref{thm cegis} gives the required conditions, finding a valid PLF that meets these requirements is challenging due to uncertainties in the system and the high dimensionality of the parameter space. Attempting an analytical solution directly may either be overly conservative or computationally impractical. To progressively fine-tune the parameters, we utilize a CEGIS framework that iteratively enhances the conditions through a process of synthesis and verification \citep{solar2005programming,jha2010oracle,abate2018counterexample}. The CEGIS framework operates through two main steps:

\smallskip
\noindent
\textbf{Synthesis Step:} This step generates a candidate PLF and polyhedral ISoA based on an initial parameterization. The parameters are chosen to meet the condition $\Psi$ in Theorem~\ref{thm cegis} using SMT solver.

\smallskip
\noindent
\textbf{Verification Step:} In this step, a SMT solver verifies whether the candidate PLF meets the condition $\Psi$ by checking the satisfiability of $\neg \Psi$ and finding a counterexample. If it does not find a counterexample, the process concludes successfully. If it does find a counterexample, it is a state  $E$ that violates the condition $\Psi$.
When a counterexample is generated, it is used to refine the parameters of the PLF. This process is repeated iteratively, ensuring that the final synthesized function adheres to all necessary conditions. 

\begin{figure}
    \centering
    \includegraphics[width=0.3\linewidth]{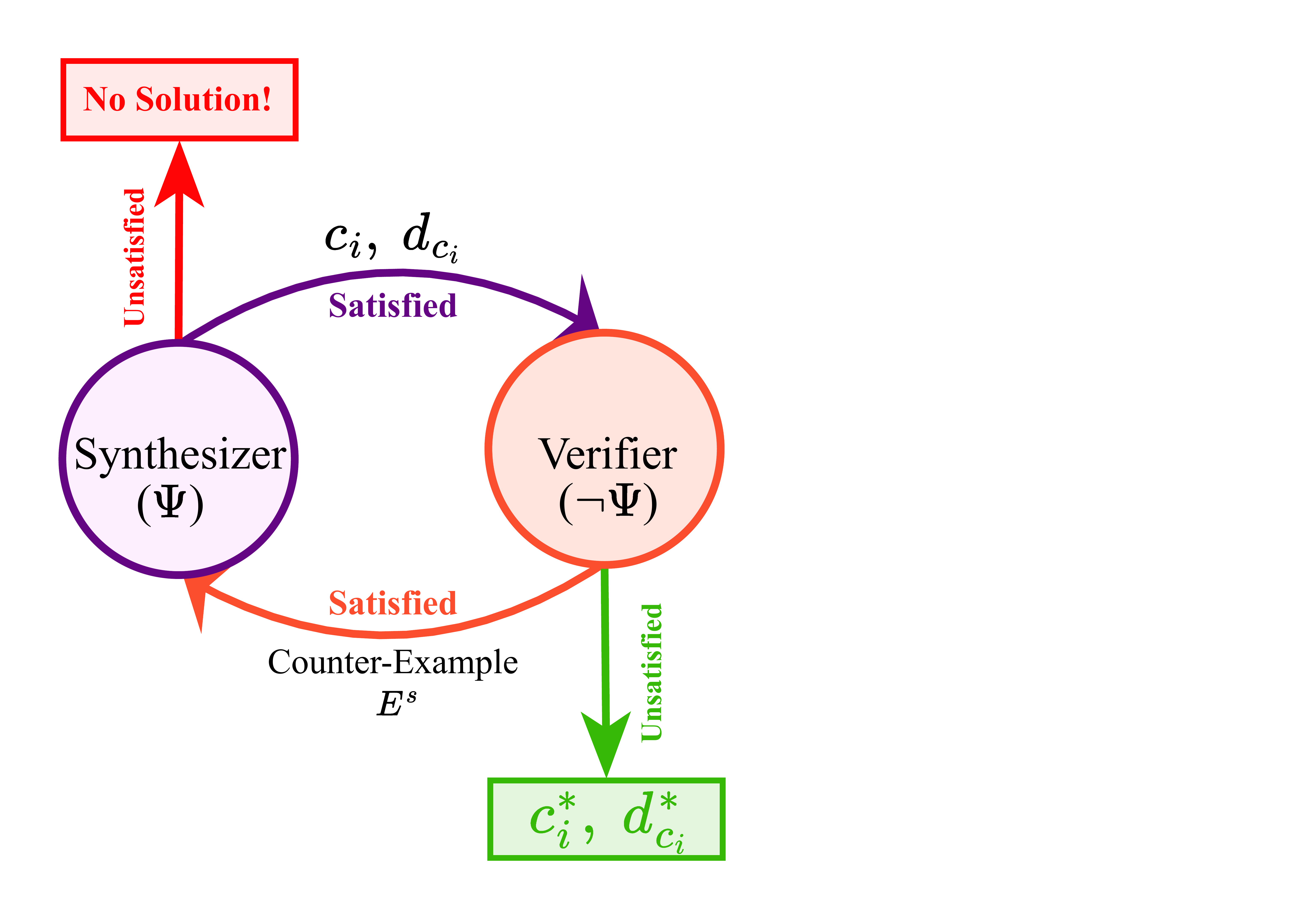}
    \caption{Representation of the CEGIS approach to find a valid PLF.}
    \label{fig: CEGIS}
\end{figure}

The structure of the CEGIS approach is illustrated in Figure~\ref{fig: CEGIS}, which shows the interaction between the synthesis step and the verification step. The complete CEGIS-based approach for designing the PLF and its polyhedral ISoA is presented in Algorithm~\ref{alg learn cegis}.

\RestyleAlgo{ruled}
\IncMargin{0.1em}
\begin{algorithm}
\hspace{3mm} 
\begin{minipage}{\dimexpr\linewidth-3mm}
\SetAlgoLined
\SetKwInOut{Input}{Input}
\SetKwInOut{Output}{Output}
\SetKwFunction{Return}{Return}
\SetKwFunction{Break}{Break}
\SetKwFunction{GenerateRandomNumber}{GenerateRandomNumber}
\SetKwFunction{Wait}{Wait}
\caption{Synthesizing PLF and its Polyhedral ISoA via SMT-based CEGIS} 
\label{alg learn cegis}
\setcounter{AlgoLine}{0}
\Input{
Matrices $A_\pi^\kappa$ and $L_\pi^\kappa$ for $\pi\in\bar{\Pi}$ and $\kappa\in\D$}
\BlankLine
\SetAlgoLined

Initialize the set of vectors $\Nu^0=\emptyset$ 

Select a finite set of states as counterexamples $\mathcal{E}^0$

\For{$s\in\N$}
{Generate $(c_i^s,d_{c_i}^s)$ for the PLF candidate $\V(E)$ using SMT solver such that $\Psi$ in~\eqref{eq pr PSI} holds for all $E\in \mathcal{E}^s$

\label{step generate} 

\If{Step~\ref{step generate} is infeasible}{
\Return \textsf{FAIL}\\
\Break}

$\Nu^{s+1} = \Nu^s \,\cup\, \{(c_i^s,d_{c_i}^s)\} $

Find a counterexample $E^s$ using condition $\neg \Psi$ in~\eqref{eq pr neg Psi} and the SMT solver

\If{No counterexample found}{
\Return $\Nu^s$ and PLF $\V(E)$\\
\Break}

$\mathcal{E}^{s+1} = \mathcal{E}^s \,\cup\,  \{E^s\}$
}

\Output{PLF $\V(E)$ with the support vectors $\Nu$}
\BlankLine
\end{minipage}
\end{algorithm}
\subsection{Synthesizing Polyhedral ISoA via Optimization-based CEGIS}\label{sec opt learn}
In this subsection, we no longer assume $\rho = 1$, as was done in the previous subsection, but we simplify the polyhedral ISoA by setting \( d_c =0\) in equation~\eqref{relax invariant set momdp}. We establish a simplified condition for the PLF and utilize optimization instead of SMT solvers. 
    \begin{theorem}
    \label{thm Phi}
    The function $\V(E)$ in \eqref{eq. poly lya} with $d_c = 0$,
the switching policy $\pi$ in \eqref{eq. switch law}, and
the ISoA $\Omega$ in \eqref{relax invariant set momdp} with $\rho>0$ give a solution for Problem~\ref{poly prblm} if the parameters of $\V(E)$ satisfy the following statement:
\begin{align}
    &\Phi = \forall E, \exists \pi,\, (\exists c_i \in \Nu, {c_i}^\top E \ge 0)\,\wedge\,\{([\forall c_i \in \Nu, (c_i^\top E) \le \rho]\,\wedge \, [\forall c_j \in \Nu,\forall \kappa \in \D, c_j^\top (A_\pi^\kappa E + L_\pi^\kappa) \leq \rho])\,\vee\,\nonumber\\
    &([\exists c_i \in \Nu,\,(c_i^\top E) > \rho]\,\wedge\,[\forall c_j \in \Nu, \forall \kappa \in \D, \exists c_\ell \in \Nu,\,(c_j^\top (A_\pi^\kappa E + L_\pi^\kappa) < c_\ell^\top E)])\}.
    \label{Thm Phi cond}
\end{align}
\end{theorem}
The proof of the theorem is relegated to the appendix.
Note that the negation of the condition $\Phi$ would be
\begin{align}
    &\neg \Phi = \exists E, \forall \pi, (\forall c_i \in \Nu,\, {c_i}^\top E < 0)\,\vee\,\{[\exists c_i \in \Nu, (c_i^\top E) > \rho] \,\,\wedge 
    [\exists c_j \in \Nu, \exists \kappa \in \D, \forall c_\ell \in \Nu,\nonumber\\
    &(c_j^\top (A_\pi^\kappa E + L_\pi^\kappa) \ge c_\ell^\top E)]\}\,\vee\,\{[\forall c_i \in \Nu,\, (c_i^\top E) \le \rho] \,\wedge [\exists c_j \in \Nu,\,\,\exists \kappa \in \D,\, c_j^\top (A_\pi^\kappa E + L_\pi^\kappa) > \rho]\}. \label{Thm neg Phi cond}
\end{align} 

We propose an optimization-based CEGIS approach that has two main steps to design the PLF and its polyhedral ISoA based on the condition $\Phi$ in~\eqref{Thm Phi cond}.

\smallskip
\noindent
\textbf{Synthesis Step:} In this step, the aim is to find the vectors $c_i\in \Nu$ based on the condition $\Phi$ in~\eqref{Thm Phi cond}. Hence, the following optimization problem is defined:
\begin{align}
     \min_{c_i \in \Nu,\rho\ge 0}& \quad \rho \nonumber \\
    & \text{s.t. } (c_i,\rho)\models \Phi,
    \label{opt learn}
\end{align}
where the optimization is constrained with the satisfaction of $\Phi$ by the vectors in $\Nu$ and $\rho\ge 0$ and selected elements for $E$.

\smallskip
\noindent
\textbf{Verification Step:}
Once the optimization problem is solved, the resulting PLF and polyhedral ISoA will be verified. Instead of checking the condition $\neg \Phi$ in~\eqref{Thm neg Phi cond} at once, the verification process is segmented into three distinct phases described in the following, with each phase focusing on a specific constraint of the condition $\neg \Phi$. 
This allows us to find potentially up to three counterexamples.
The complete process is detailed in Algorithm~\ref{alg learn opt}.

Below, we define three optimization problems corresponding to each of the constraints in $\neg \Phi$ with $E_1 = [e_{1}^1,e_{1}^2,\ldots,e_{1}^{n \times q}]^\top$, $E_2 = [e_{2}^2,e_{2}^2,\ldots,e_{2}^{n \times q}]^\top$, and $E_3 = [e_{3}^1,e_{3}^2,\ldots,e_{3}^{n \times q}]^\top$ as follows:
\begin{align}\label{eq phase 1}
    \max&_{E_1}\left\{ \Sigma_{k=1}^{n\times q}e_{1}^k \text{ s.t. } \forall c_i \in \Nu,\, {c_i}^\top E < 0\right\},\\
     \max&_{E_2}\{ \Sigma_{k=1}^{n\times q}e_{2}^k  \text{ s.t. } \forall \pi,\, [\exists c_i \in \Nu, (c_i^\top E) > \rho] \,\,\wedge 
    [\exists c_j \in \Nu, \exists \kappa \in \D, \forall c_\ell \in \Nu, (c_j^\top (A_\pi^\kappa E + L_\pi^\kappa) \ge c_\ell^\top E)]\},\label{eq phase 2}\\
    \max&_{E_3}\{ \Sigma_{k=1}^{n\times q}e_{3}^k \text{ s.t. }\forall \pi, [\forall c_i \in \Nu,\, (c_i^\top E) \le \rho] \,\wedge\, [\exists c_j \in \Nu,\,\,\exists \kappa \in \D,  c_j^\top (A_\pi^\kappa E + L_\pi^\kappa) > \rho]\}. \label{eq phase 3}
\end{align}
Since the sets \(\Nu\), \(\D\), and \(\mathcal{E}\) are finite, the quantifiers and disjunctions involved in the synthesis and verification steps can be translated into a mixed-integer linear programming (MILP) optimization. Universal quantifiers are represented by repeating constraints over their respective index sets, while existential quantifiers and disjunctions are captured using binary selection or activation variables through big-M or indicator constraints~\citep{williams2013model, conforti2014integer, vielma2015mixed}. This approach allows for the problem to be solved directly using MILP solvers such as $\mathsf{MOSEK}$ or $\mathsf{Gurobi}$~\citep{mosek-matlab-toolbox, gurobi}. Modeling tools like $\mathsf{MOSEK\,Fusion}$ and $\mathsf{Pyomo}$ (via $\mathsf{pyomo.gdp}$) directly support disjunctions~\citep{mosekfusion, hart2011pyomo, bynum2021pyomo}.

\RestyleAlgo{ruled}
\IncMargin{0.1em}
\begin{algorithm}
\hspace{3mm} 
\begin{minipage}{\dimexpr\linewidth-3mm}
\SetAlgoLined
\SetKwInOut{Input}{Input}
\SetKwInOut{Output}{Output}
\SetKwFunction{Return}{Return}
\SetKwFunction{Break}{Break}
\SetKwFunction{GenerateRandomNumber}{GenerateRandomNumber}
\SetKwFunction{Wait}{Wait}
\caption{
Synthesizing PLF and its Polyhedral ISoA via Optimization-based CEGIS} 
\label{alg learn opt}
\setcounter{AlgoLine}{0}
\Input{
Matrices $A_\pi^\kappa$ and $L_\pi^\kappa$ for $\pi\in\bar{\Pi}$ and $\kappa\in\D$}
\BlankLine
\SetAlgoLined
Initialize the set of vectors $\Nu^0=\emptyset$

Select a finite set of states as counterexamples $\mathcal{E}^0$

\For{$s\in\N$}
{Generate $c_i^s$ for the PLF candidate $\V(E)$ with $d_c = 0$ by solving the optimization~\eqref{opt learn}
with constraints evaluated on $E \in \mathcal{E}^s$
\label{stp1}

\If{Step~\ref{stp1} is infeasible}{
\Return \textsf{FAIL}\\
\Break}

$\Nu^{s+1} = \Nu^s \,\cup\, \{c_i^s\}$

Set flags $\delta_1 = 0, \delta_2 = 0, \delta_3 = 0$

Find $E_1^s$ by solving the optimization~\eqref{eq phase 1}
\label{stp ph1}

\If{Step~\ref{stp ph1} is infeasible}{$\delta_1 = 1$}

Find $E_2^s$ by solving the optimization~\eqref{eq phase 2}
\label{stp ph2}

\If{Step~\ref{stp ph2} is infeasible}{$\delta_2 = 1$}

Find $E_3^s$ by solving the optimization~\eqref{eq phase 3}\label{stp ph3}

\If{Step~\ref{stp ph3} is infeasible}{ $\delta_3 = 1$}

\If{$\delta_1\delta_2\delta_3 == 1$}{
\Return $\Nu^s$ and the PLF $\V(E)$\\
\Break}

$\mathcal{E}^{s+1} = \mathcal{E}^s \cup  \{E_1^s,E_2^s,E_3^s\}$
}

\Output{PLF $\V(E)$ with the support vectors $\Nu$}
\BlankLine
\end{minipage}
\end{algorithm}
\subsection{Robust Lyapunov-Based VI Algorithm} 
\label{sec: robust vi alg}
Both Algorithms~\ref{alg learn cegis} and \ref{alg learn opt} give two different solutions to Problem~\ref{poly prblm} for computing a PLF and its polyhedral ISoA. By using these algorithms, the PLF $\V$ and polyhedral ISoA $\Omega$ can be found for error dynamics in~\eqref{Augmented E moimdp}. Building on these results, we propose a systematic algorithm that integrates the Lyapunov-based VI procedure, outlined in Algorithm~\ref{relax vi alg moimdp}. For a general $W_{tar}$ not in $\f$, the algorithm first finds a $W_{tar}'\in\f$ that is closest to $W_{tar}$, then applies the results of Theorems~\ref{thm cegis} and \ref{thm Phi} to compute the policy. The computed invariant set $\Omega$ is then shifted with $W_{tar}'$ and, if needed, expanded to include $W_{tar}$, thus giving $\g$ as a solution to Problem~\ref{moimdp prblm}.
\RestyleAlgo{ruled}
\IncMargin{0.1em}
\begin{algorithm}
\hspace{3mm} 
\begin{minipage}{\dimexpr\linewidth-3mm}
\SetAlgoLined
\SetKwInOut{Input}{Input}
\SetKwInOut{Output}{Output}
\SetKwFunction{GenerateRandomNumber}{GenerateRandomNumber}
\SetKwFunction{Wait}{Wait}
\caption{Robust Lyapunov-Based VI Algorithm for MOIMDP} 
\label{relax vi alg moimdp}
\setcounter{AlgoLine}{0}
\Input{
  MOIMDP \( \Sigma = (X, x_0, U, \underbar{P}, \overbar{P}, \underbar{R}, \overbar{R}) \) and target values $W_{tar}$
}
\BlankLine
\SetAlgoLined
Determine the set of stationary policies $\bar{\Pi}$

Determine $A_\pi^j$ and $L_\pi^j$ for all $(\pi,j) \in \bar{\Pi} \times \D$ according to~\eqref{politopic uncertainity moimdp}

Compute $\Ad_\pi$, $\hat{B}_\pi$ for all $\pi \in\bar{\Pi}$ according to~\eqref{nominal matrices}

Define $\f$ as in~\eqref{feasable set imdp}

Compute $W_{tar}' := \argmin_{W'} \{\|W' - W\|, W\in\f\}$

Select $\lambda \in \Lambda_{M}$ with $W_{tar}' = (I-\Ad_{\lambda})^{-1}\hat{B}_{\lambda}$ and compute $\Ad_{\lambda}$ and $\hat{B}_{\lambda}$ according to~\eqref{Aland Blanda imdp}

Compute PLF $\V(E)$ and polyhedral ISoA $\Omega$ via Algorithm~\ref{alg learn cegis} or Algorithm~\ref{alg learn opt}
  
Compute $\g$ as the smallest set such that $\Omega + W_{tar}'\subset \g$ and $W_{tar}\in\g$
  
Set $W_0 = 0 $ and $E_0=W_0-W_{tar}'$

\For{$k\in\N$}
{Compute ${\pi}_k(E_k)$ using~\eqref{eq. switch law}\;

Update $E_{k+1}$ using~\eqref{Augmented E moimdp} and by applying $\textsf{opt}$ to uncertainty
}

\Output{Set $\g$ and switching policy $ \boldsymbol{\pi} = (\pi_0,\pi_1,\ldots)$ such that $W_{tar}\in\g$, $W_{\boldsymbol{\pi},\textsf{opt}}\in\g$
}
\BlankLine
\end{minipage}
\end{algorithm}
\subsection{Computational Complexities}
\noindent\textbf{Computational Complexity of Algorithm~\ref{alg learn cegis}.}
The SMT solver $\mathsf{Z3/Z3py}$ does not offer closed-form worst-case complexity guarantees for quantified SMT. The tool employs the DPLL(T) algorithm, which combines a SAT search over a Boolean structure with polynomial-time checks for linear real arithmetic (LRA). As a result, the performance of $\mathsf{Z3/Z3py}$ is dependent on the specific instance being solved. Moreover, even in the absence of quantifiers, SMT involving linear inequalities can become NP-complete once disjunctions are introduced. While the checks for linear programming are polynomial in time, the Boolean search can be exponential. Considering $|\mathcal{E}| = |\mathcal{\Nu}|= s$ at step \( s \) in the synthesizer, we will have \( 4|\bar{\Pi}| s^3 \) Boolean searches, accompanied by \( (nq + d)s^2 + s \) LRAs. As a result, the number of policies will increase the complexity exponentially, while the number of states, objectives, and uncertainty vertices will lead to a polynomial increase in complexity. Similarly, at step \( s \) in the verifier, we will encounter \( s^4 |\mathbb{D}| (1 + |\mathbb{D}|) \) Boolean searches, along with \( |\bar{\Pi}| \times \max(s + 1, nq) \) LRAs. Consequently, the number of uncertainty vertices will increase the complexity exponentially, and the number of states, objectives, and policies will contribute to a polynomial increase in complexity in the verifier.

\smallskip
\noindent\textbf{Computational Complexity of Algorithm~\ref{alg learn opt}.}
Disjunctive constraints (DJC) are often reformulated as mixed-integer problems for optimization. Software like $\mathsf{MOSEK}$ attempts to replace DJCs with big-M constraints, simplifying them in the process. The computational effort to solve mixed-integer problems grows exponentially with the size, making them NP-hard. For instance, a problem with \( n \) binary variables may require \( 2^n \) relaxations (see \citep{mosek-matlab-toolbox,mosekfusion}). In our setting, the synthesizer utilizes optimization with DJCs. Translating these into MILP at step \( s \) results in \( (|\bar{\Pi}| + 3s + 1) \) integer variables, causing exponential complexity growth with the number of policies. The verifier operates in three phases: (i) Phase 1 in \eqref{eq phase 1} can be translated into an LP which is solvable in polynomial time.
(ii) Phase 2 in \eqref{eq phase 2} has DJCs. When converted to MILP, it has \( s^2 |\mathbb{D}| \) integer variables, leading to exponential complexity due to the vertices in the uncertainty set. 
(iii) Phase 3 in \eqref{eq phase 3} also has DJCs, resulting in \( s |\mathbb{D}|\) integer variables and resulting in exponential complexity due to the vertices in the uncertainty set. Thus, the complexity in Algorithm~\ref{alg learn opt} increases significantly with the number of policies and uncertainty set vertices.

\section{Case Studies}\label{sec: case studies}
We apply Algorithm~\ref{relax vi alg moimdp} on three case studies. The first one is the MDP model of a recycling robot \citep{sutton2018reinforcement, doi:10.1080/00207179.2021.2005260}. The second one is an IMDP model adopted from \citep{hahn2019interval, monir2024switch}. The third case study is an IMDP model for the life cycle of the battery of an electric vehicle (EV), adopted from~\citep{thein2014decision}.

\subsection{Recycling Robot with an MDP Model}
\label{ex mdp}
The transition probability matrices in the model of the recycling robot $P(\pi)$ for $\pi\in\{1,\ldots,6\}$ are
\begin{align*}
    &P(1)=\begin{bmatrix}
    \beta & 1-\beta \\
    1-\alpha & \alpha
    \end{bmatrix},
    \quad P(2)=\begin{bmatrix}
    \beta & 1-\beta \\
    1 & 0
    \end{bmatrix}, \quad P(3)= \begin{bmatrix}
    1 & 0 \\
    1-\alpha & \alpha
    \end{bmatrix},\quad
    P(4)=\begin{bmatrix}
    1 & 0 \\
    0 & 1
    \end{bmatrix}, \\
    &P(5)=\begin{bmatrix}
    0 & 1 \\
    1-\alpha & \alpha
    \end{bmatrix}, \quad P(6)=\begin{bmatrix}
    0 & 1 \\
    0 & 1
    \end{bmatrix},
\end{align*}
with $\alpha=0.7$ and $\beta=0.4$. The reward vectors are $R(1) =[
        \beta r_{search} - 3 (1-\beta) ;
        r_{search}
    ]$, $R(2) = [
        \beta r_{search} - 3 (1-\beta) ;
        r_{wait}
    ]$, $R(3) =[
        r_{wait} ;
        r_{search}
    ]$, $R(4) = [
        r_{wait} ;
        r_{wait}
    ]$, $R(5) = [
        0 ;
        r_{search}
    ]$, and $R(6) = [
        0 ;
        r_{wait}
    ]$, with $r_{search}=8$ and $r_{wait}=2$. The discount factor $\gamma$ has been set to 0.5. 
We use $W_{tar}$ corresponding to
$\lambda = {[0;0;1;0;0;0]}$ and initial values $W_0 = {[0;0]}$. 

\begin{figure}
    \centering
    \begin{minipage}[b]{0.16\linewidth}
        \includegraphics[width=\linewidth]{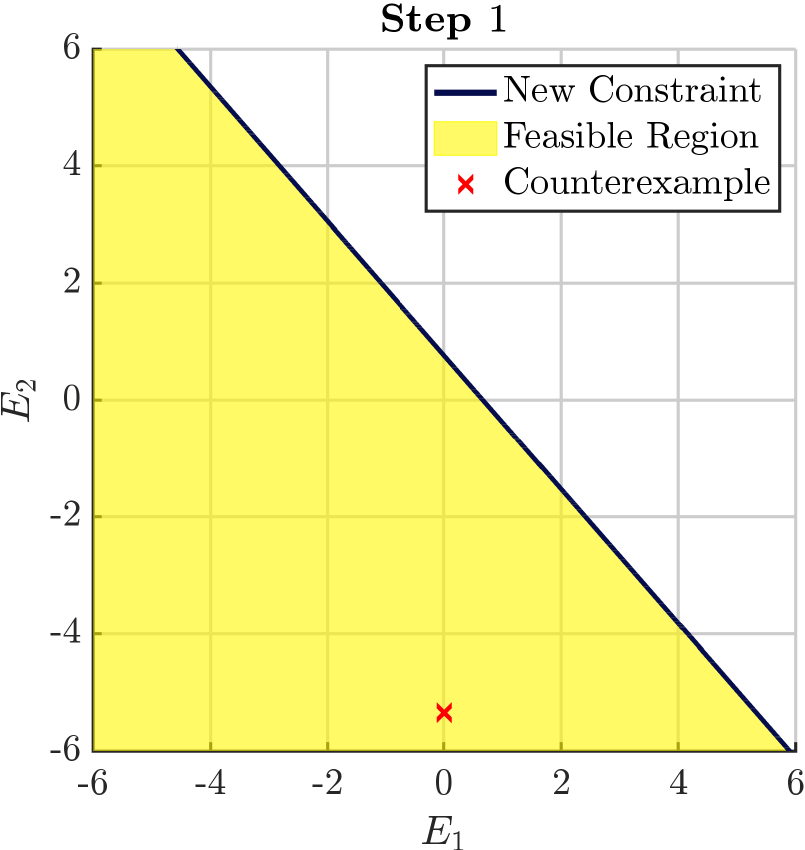}
        \centering (a)
    \end{minipage}
    \begin{minipage}[b]{0.16\linewidth}
        \includegraphics[width=\linewidth]{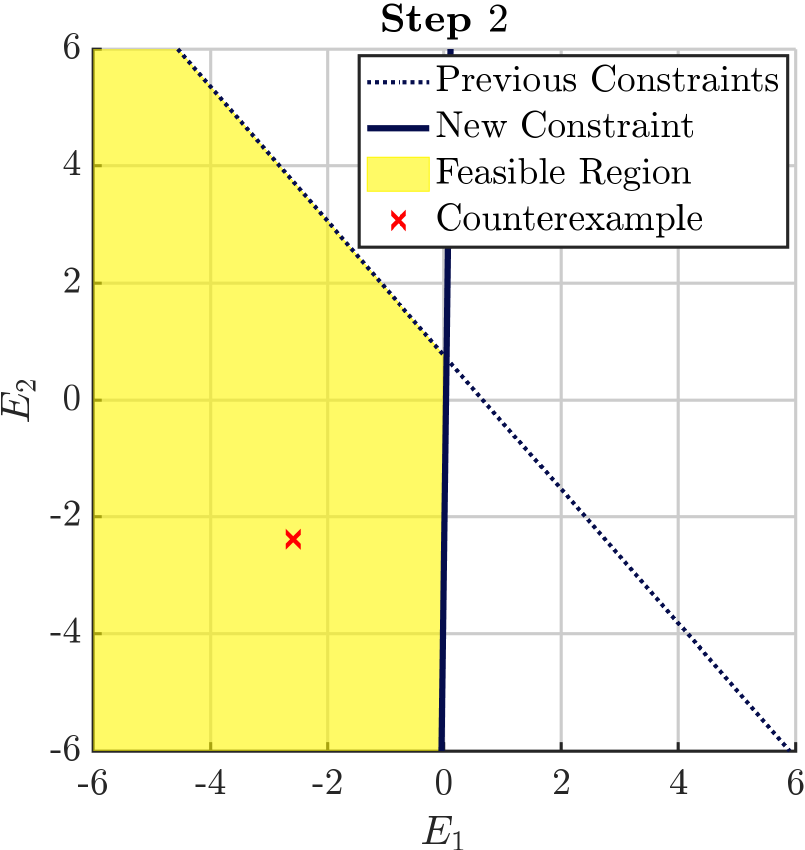}
        \centering (b)
    \end{minipage}
    \begin{minipage}[b]{0.16\linewidth}
        \includegraphics[width=\linewidth]{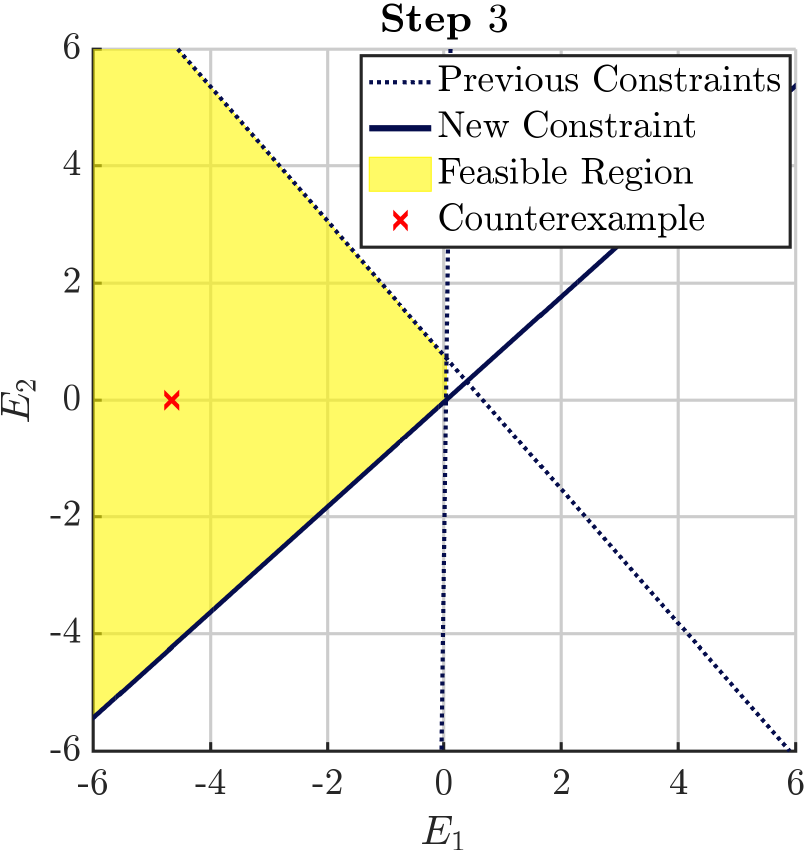}
        \centering (c)
    \end{minipage}
    \begin{minipage}[b]{0.16\linewidth}
        \includegraphics[width=\linewidth]{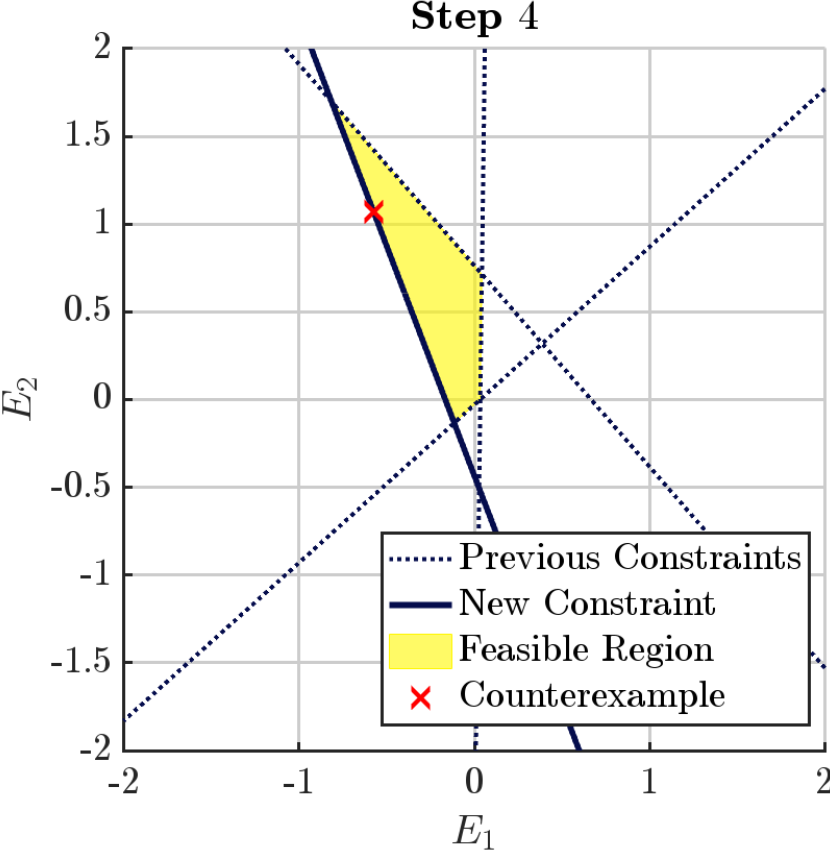}
        \centering (d)
    \end{minipage}
    \begin{minipage}[b]{0.16\linewidth}
        \includegraphics[width=\linewidth]{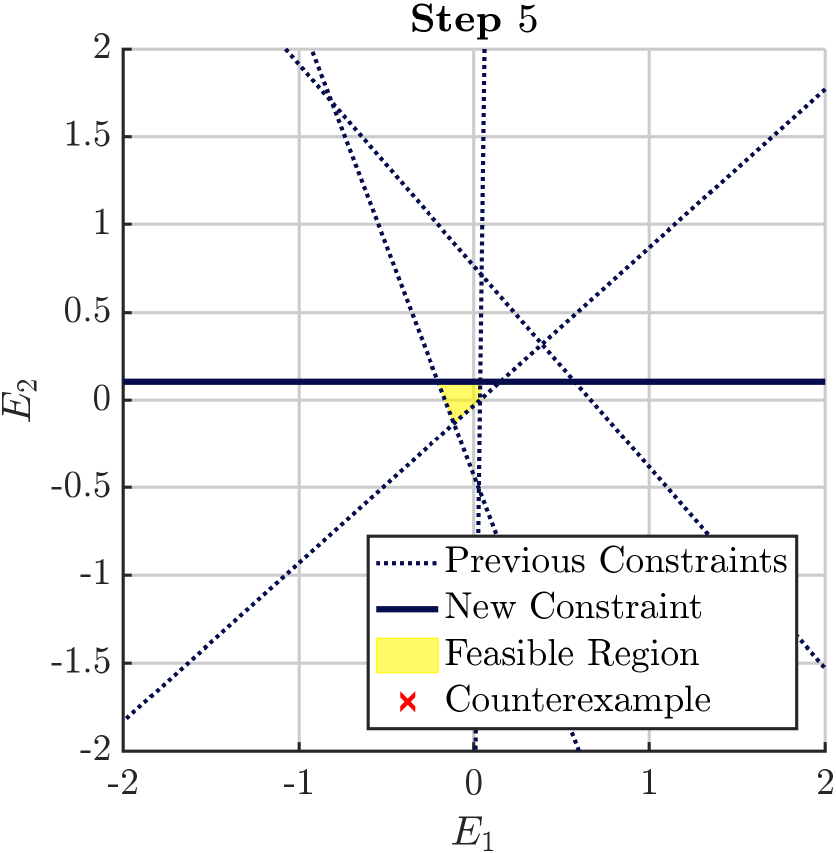}
        \centering (e)
    \end{minipage}
    \begin{minipage}[b]{0.16\linewidth}
        \includegraphics[width=\linewidth]{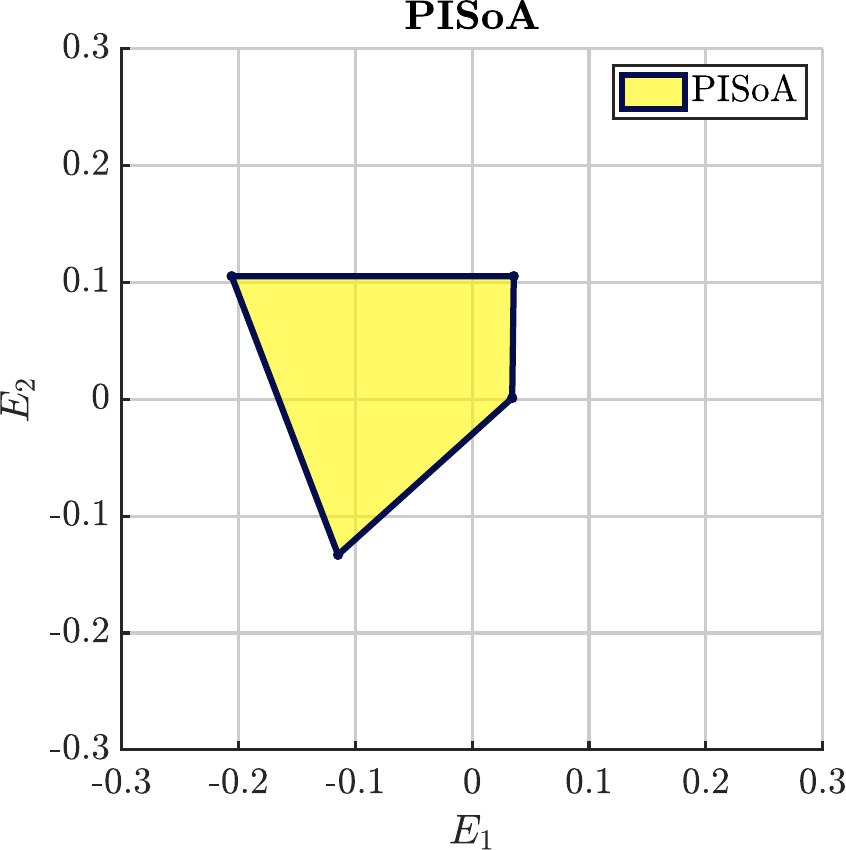}
        \centering (f)
    \end{minipage}
    \caption{\textbf{Recycling Robot.} This illustration depicts the CEGIS procedure of Algorithm~\ref{alg learn cegis} for computing the polyhedral ISoA. (a) The learner proposes an initial PLF candidate that creates a feasible region (yellow), but the verifier identifies a counterexample (red cross). (b)–(d) Each counterexample generates a new constraint that refines the feasible region, and the updated candidate is re-evaluated by the verifier. (e) After five iterations, no further counterexamples are found, indicating that the candidate PLF satisfies all the required conditions. (f) The final polyhedral ISoA is defined by the resulting feasible region.}
    \label{fig: steps}
\end{figure}

\begin{figure}
    \centering
    \begin{minipage}[b]{.33\linewidth}
        \includegraphics[width=\linewidth]{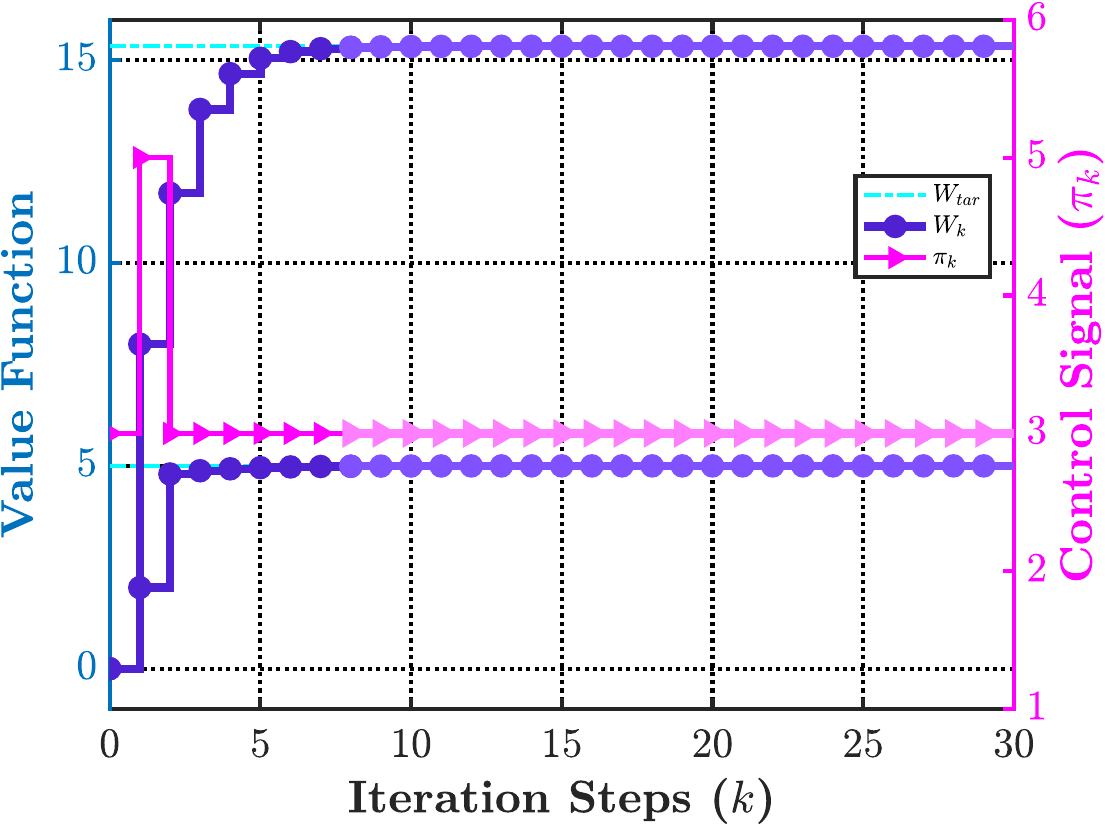}
        \centering (a)
    \end{minipage}
    \begin{minipage}[b]{.31\linewidth}
        \includegraphics[width=\linewidth]{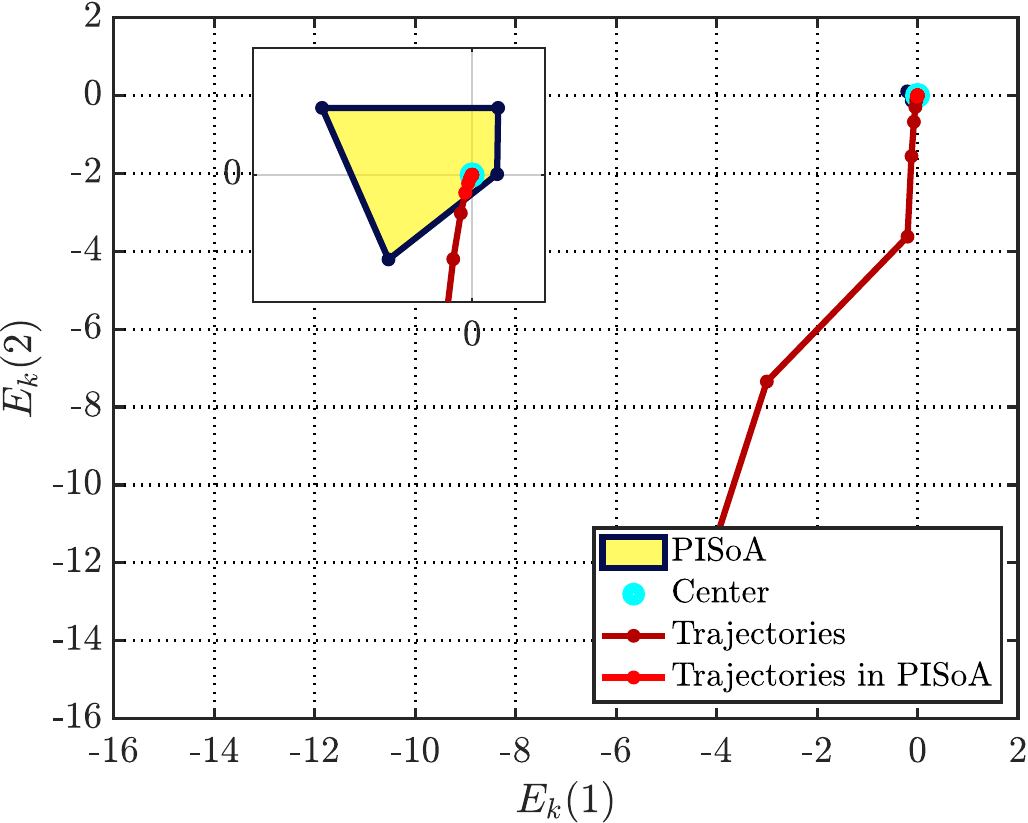}
        \centering (b)
    \end{minipage}
    \begin{minipage}[b]{.315\linewidth}
        \includegraphics[width=\linewidth]{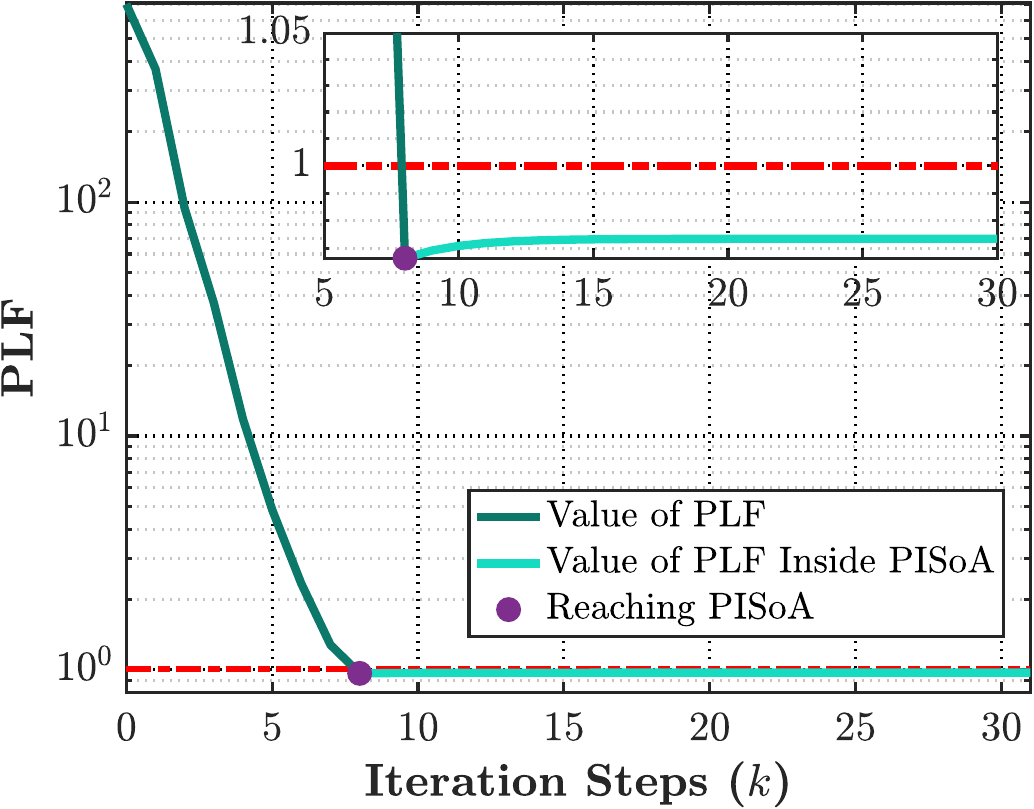}
        \centering (c)
    \end{minipage}
    \caption{\textbf{Recycling Robot.} Results of Algorithm~\ref{relax vi alg moimdp}. (a) Iterative evolution of the value function $W_k$, which converges to $W_{\text{tar}}$, together with the synthesized policy $\pi_k$. (b) Error trajectories $E_k$ under the computed policy, all converging to the polyhedral ISoA and remaining inside thereafter. (c) Evolution of the PLF, which decreases monotonically, drops below $1$, and stays under $1$, confirming invariance.}
    \label{fig: mdp alg1}
\end{figure}

We apply Algorithm~\ref{relax vi alg moimdp} with Algorithm~\ref{alg learn cegis} as its subroutine to compute the policy for this model. 
The intermediate steps of the CEGIS synthesis process are illustrated in Fig.~\ref{fig: steps}.
After computing the polyhedral ISoA, we continue with the remaining steps of Algorithm~\ref{relax vi alg moimdp} to compute the policy. Figure~\ref{fig: mdp alg1} illustrates the evolution of the value function, the resulting policy, the convergence of the error trajectories towards the computed polyhedral ISoA, and the decrease of the PLF. These results show that the value iteration converges, the trajectories stay within the polyhedral ISoA, and the PLF decreases below $\rho=1$, then remains under this value, confirming invariance.

\subsection{An IMDP Example}
\label{ex imdp}
\begin{figure}
    \centering\includegraphics[width=.4\linewidth]{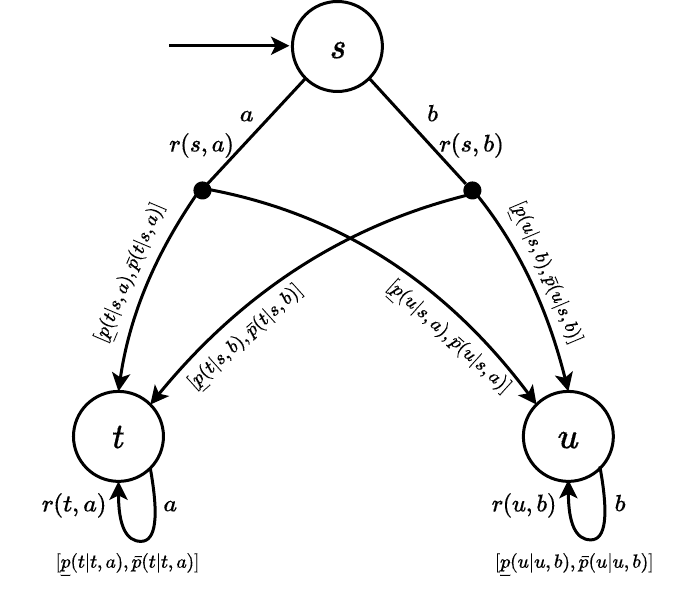}
    \caption{IMDP Example adopted from \citep{hahn2019interval}.}
    \label{fig: imdp ex}
\end{figure}

    {Figure~\ref{fig: imdp ex} shows an IMDP model adopted from \citep{hahn2019interval}}, with the set of states $X = \{s, t, u\}$, the initial state $s$, and the set of actions $U = \{a,b\}$. The non-zero transition probability intervals are illustrated in the graph. The set of policies is $\bar\Pi = \{1,2\}$. The interval transition probability matrices are defined as
\begin{align*}
    \underbar{P}(1) = \begin{bmatrix}
    0 & \frac{1}{3} & \frac{1}{10} \\ 
    0 & 1 & 0 \\ 
    0 & 0 & 1
\end{bmatrix}, \quad
\overbar{P}(1) = \begin{bmatrix}
    0 & \frac{2}{3} & 1 \\ 
    0 & 1 & 0 \\ 
    0 & 0 & 1
\end{bmatrix}, \quad
\underbar{P}(2) = \begin{bmatrix}
    0 & \frac{2}{5} & \frac{1}{10} \\ 
    0 & 1 & 0 \\ 
    0 & 0 & 1
\end{bmatrix}, \quad 
\overbar{P}(2) = \begin{bmatrix}
    0 & \frac{3}{5} & 1 \\ 
    0 & 1 & 0 \\ 
    0 & 0 & 1
\end{bmatrix}.
\end{align*}
The interval reward vectors are defined as $\underbar{R}(1) = \left[\frac{13}{10}; \frac{3}{10};
        \frac{1}{10}\right]$, $\overbar{R}(1) = \left[5; \frac{3}{10};
        \frac{1}{10}\right]$, $\underbar{R}(2) = \left[\frac{1}{2}; \frac{3}{10};
        \frac{1}{10}\right]$, and $\overbar{R}(2) = \left[\frac{8}{5}; \frac{3}{10};
        \frac{1}{10}\right]$. The discount factor $\gamma = 0.7$ 
        and target $W_{tar}$ corresponding to $\lambda = {\left[\frac{9}{10};\frac{1}{10}\right]}$ are selected with the initial values $W_0={\left[0;0;0\right]}$.

\begin{figure}
    \centering
    \includegraphics[width=0.5\linewidth]{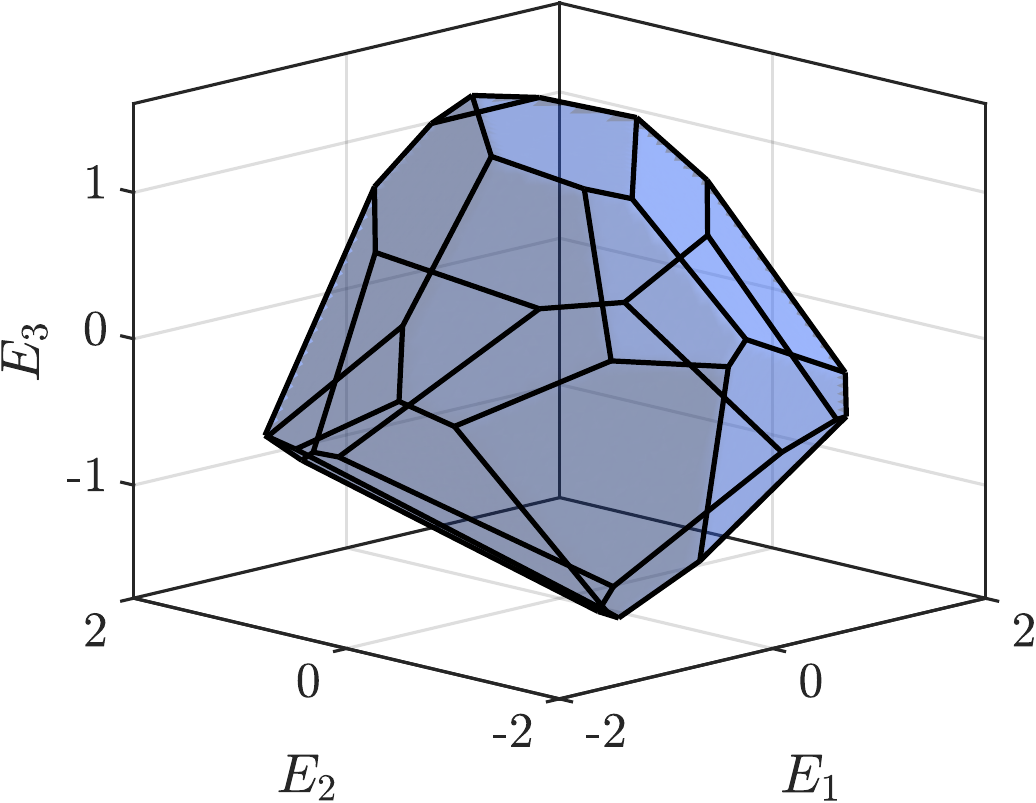}
    \caption{\textbf{IMDP Example.} Polyhedral ISoA obtained for the error dynamics of the IMDP example using Algorithm~\ref{alg learn opt} within Algorithm~\ref{relax vi alg moimdp}.}
    \label{fig imdp pisoa}
    \end{figure}

\begin{figure}
    \centering
    \begin{minipage}[b]{.33\linewidth}
        \includegraphics[width=\linewidth]{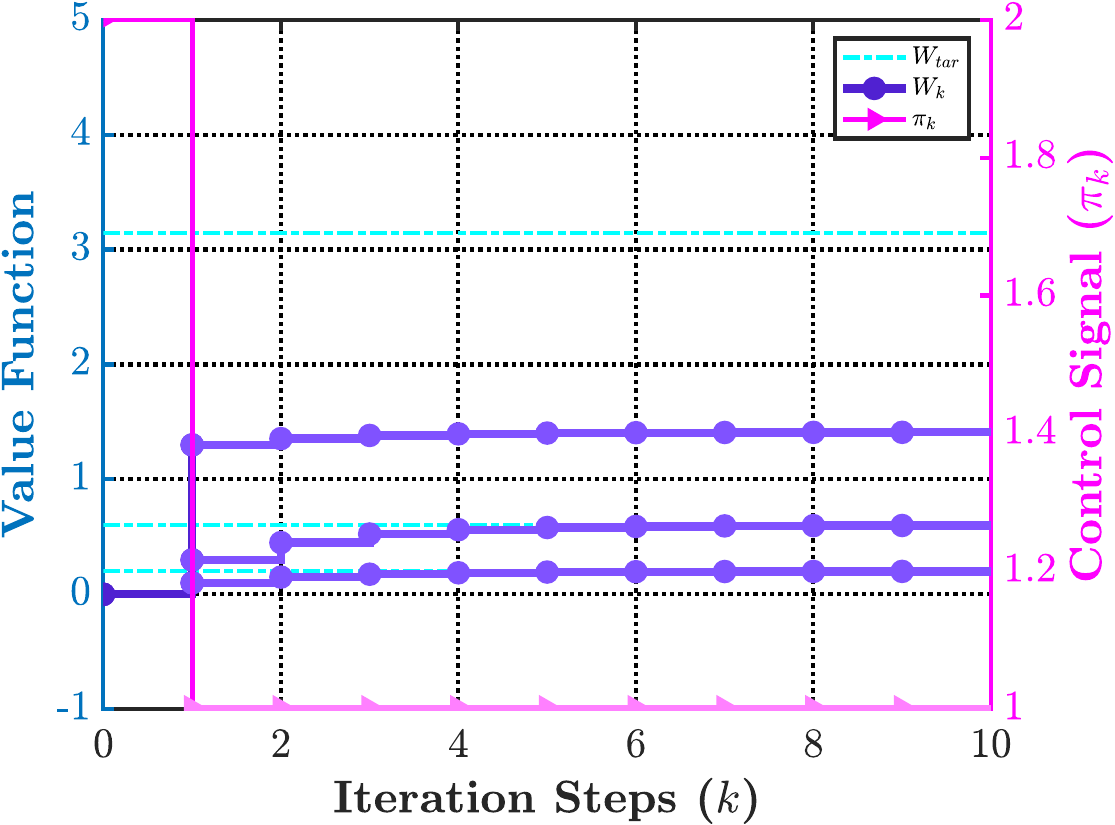}
        \centering (a)
    \end{minipage}
    \begin{minipage}[b]{.33\linewidth}
        \includegraphics[width=\linewidth]{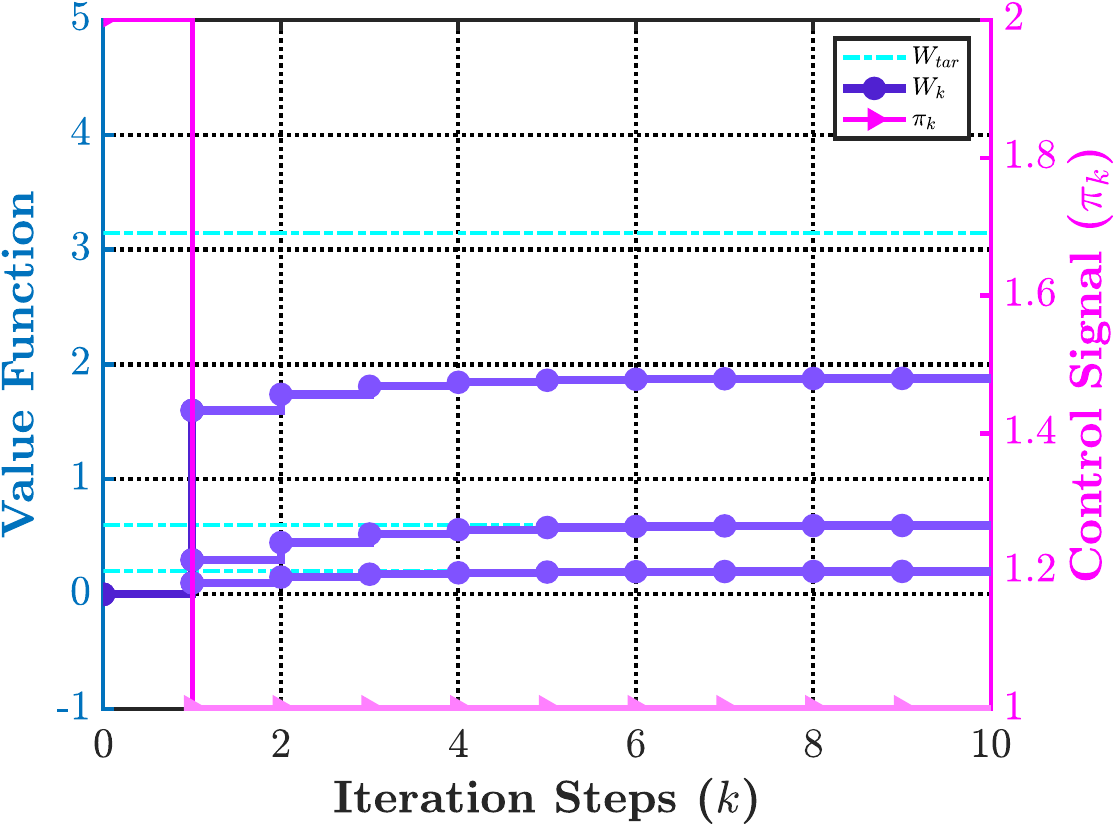}
        \centering (b)
    \end{minipage}
    \begin{minipage}[b]{.315\linewidth}
        \includegraphics[width=\linewidth]{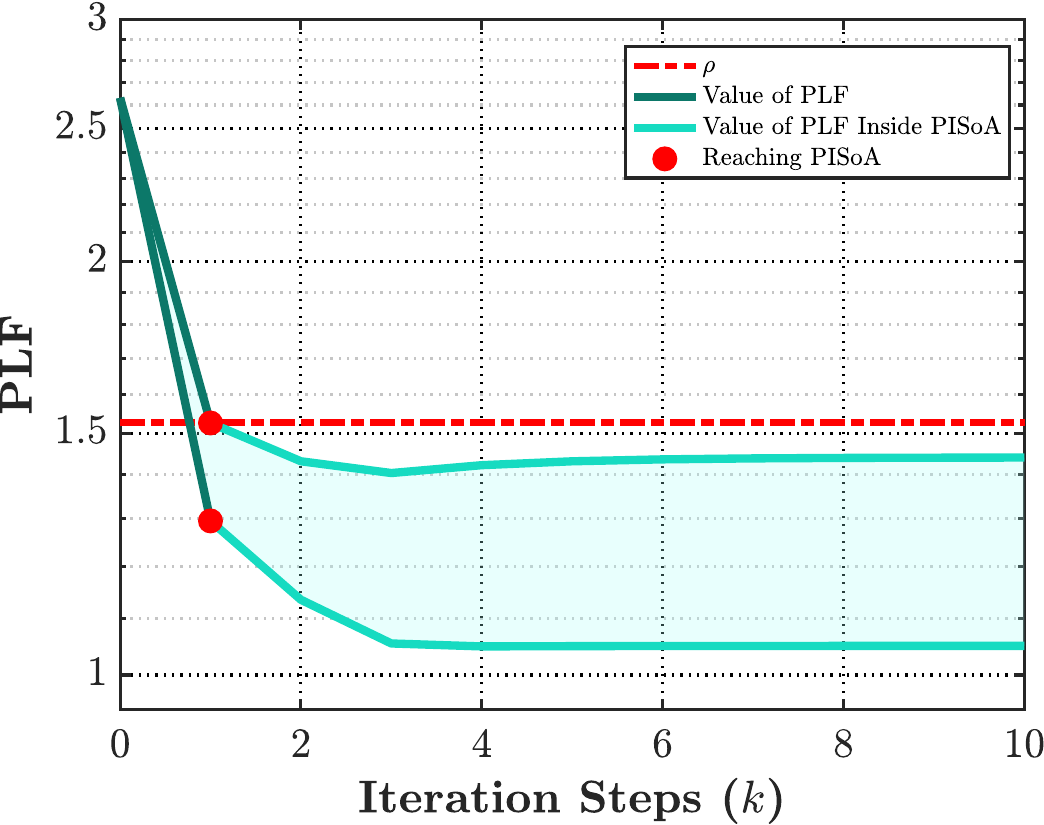}
        \centering (c)
    \end{minipage}
    \caption{\textbf{IMDP Example.} Results of Algorithm~\ref{relax vi alg moimdp}. (a) and (b) The evolution of the value functions and selected policies for both the lower bound and upper bound of the objective function. (c) The PLF for these bounds decreases below $\rho$ and remains within the invariant set.}
    \label{fig:imdp values}
\end{figure}

We apply Algorithm~\ref{relax vi alg moimdp} with Algorithm~\ref{alg learn opt} as its subroutine to compute the policy for this model.
The obtained polyhedral ISoA is shown in Figure~\ref{fig imdp pisoa}. The value functions for the lower and upper bound of the objective function together with the switching strategy is shown in Figure~\ref{fig:imdp values}. the figure also shows that the computed PLF decreases
below $\rho$, then remains under $\rho$, confirming invariance

\subsection{Life Cycle of an EV Battery}
\label{subsec:ev-mdp}
\begin{figure}
    \centering
\includegraphics[width=.4\linewidth]{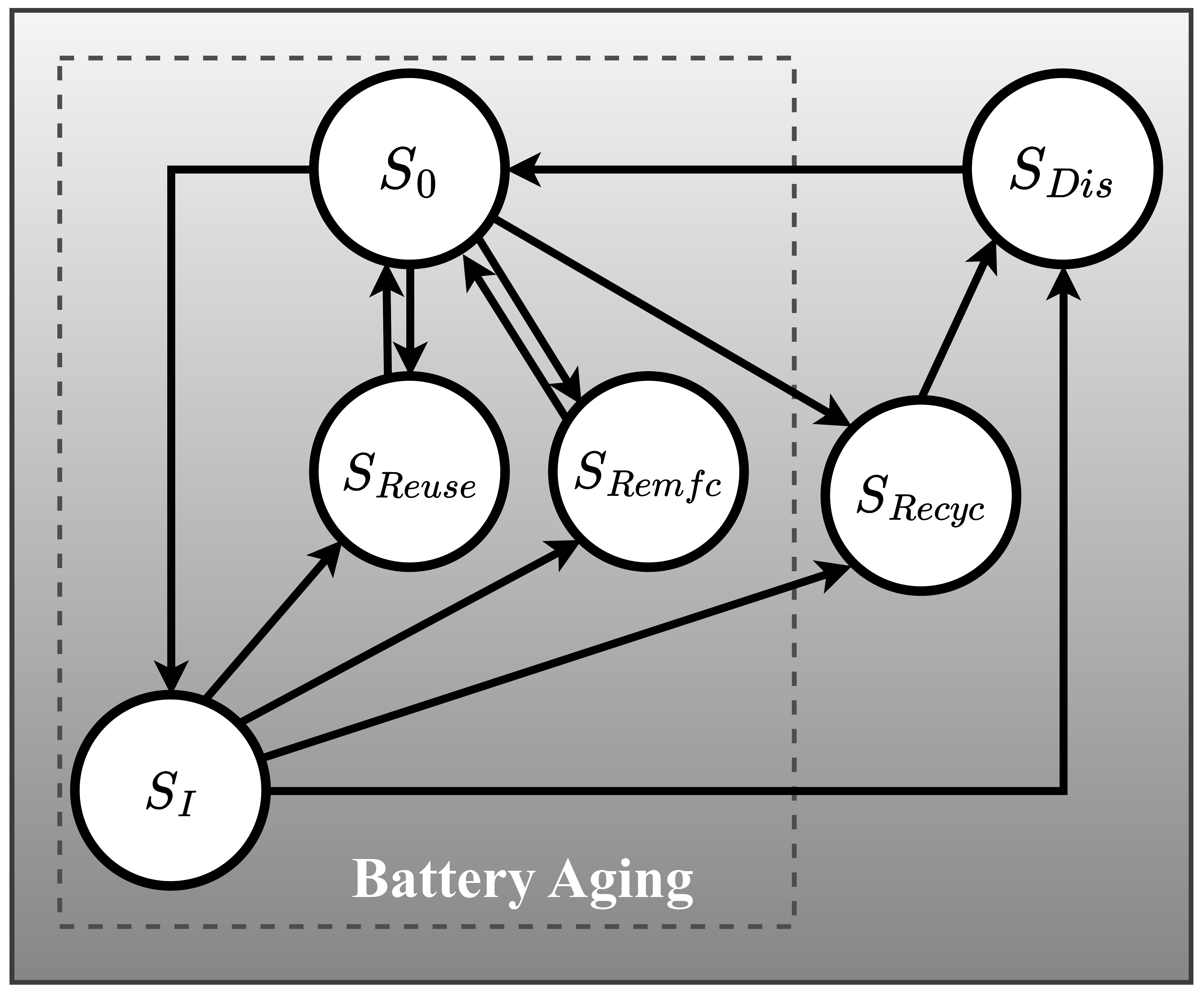}
    \caption{MDP model for the life cycle of the EV battery.}
    \label{fig:ev decision model}
\end{figure}

We consider the lithium-ion EV battery life cycle model adopted from \citep{thein2014decision}. We model the end-of-life routing problem for EV batteries as a discounted IMDP illustrated in Figure~\ref{fig:ev decision model}. The model has six states: \(X = \{S_0, S_I, S_R, S_M, S_C, S_D\}\). These states correspond to the following stages: \emph{aging in use} \((S_0)\), \emph{inspection} \((S_I)\), \emph{reuse} \((S_R)\), \emph{remanufacture} \((S_M)\), \emph{recycling} \((S_C)\), and \emph{disposal} \((S_D)\). Decisions can only be made at the decision states \(\{S_0, S_I\}\), where the available actions in state $S_0$ are \(\mathcal{A}_0 = \{\text{Inspect}, \text{Reuse}, \text{Remanufacture}, \text{Recycle}\}\) and in state $S_I$ are \(\mathcal{A}_I = \{\text{Reuse}, \text{Remanufacture}, \text{Recycle}, \text{Dispose}\}\). In the remaining states \(\{S_R, S_M, S_C, S_D\}\), no decisions are made. To account for modeling errors and operational variability, the nonzero entries of the transition probability matrix have up to \(3\%\) variation.

The rewards are assigned for each action and state, categorized into three types: (i) \emph{Economic cost}: Every action involves a nominal cost, reflecting the effort and resources expended (e.g., remanufacturing is more costly than reusing; recycling and disposal incur handling and compliance expenses). These costs serve as negative rewards and remain constant for each specific action. (ii) \emph{Health penalty}: this captures expected wear on the battery; actions that keep or return batteries to service (aging/reuse) incur higher penalties than facility-based processing steps (remanufacture, recycle, disposal). We allow slight adjustments in each state to accurately reflect where the action is applied. (iii) \emph{Environmental benefit}: circular actions get positive rewards (reuse $>$ remanufacture $>$ recycle), while disposal is negative. Each reward uses nominal action-based values with symmetric relative intervals ($0.01–0.03\%$ to reflect estimation error).

We apply three implementations: (i) \algCE, which instantiates Algorithm~\ref{relax vi alg moimdp} using the SMT-based CEGIS of Algorithm~\ref{alg learn cegis}; (ii) \algOP, which instantiates Algorithm~\ref{relax vi alg moimdp} using the optimization-based CEGIS of Algorithm~\ref{alg learn opt}; and (iii) \algEL, which adapts the quadratic Lyapunov function from the baseline paper~\citep{monir2024lyapunov}.
We consider three scenarios—single-objective, bi-objective, and tri-objective—to evaluate the effects of increasing objective dimensions. The quantitative outcomes are summarized in Table~\ref{tab:ev-results} that include the convergence behavior, lower and upper bounds under interval uncertainty, and computation time. 
\begin{table}
  \centering
  \caption{\textbf{Life Cycle of an EV Battery.}
  Number of states is $n=6$ and the number of objectives is \(q\). The computed policy is $\pi$.
  For each number of objectives \(q \in \{1, 2, 3\}\), three target vectors $W_{\mathrm{tar}}$ are considered (rows \(W_1\) to \(W_9\)).
  The norms of the lower and upper bounds of the error are reported in the columns \(\lVert \underbar{E}\rVert,\lVert \overbar{E}\rVert\).
  The computational times are 
  \(\mathcal{T}_1\) for PLF synthesis in Algorithm~\ref{alg learn cegis},
  \(\mathcal{T}_2\) for PLF synthesis in Algorithm~\ref{alg learn opt}, and
  \(\mathcal{T}_3\) for value iteration of Algorithm~\ref{relax vi alg moimdp}. The computational time of the baseline algorithm is \(\mathcal{T}\), which gives a quadratic function.
  In row $W_6$, the `--' entries in the \algEL block indicate that the baseline optimization problem is infeasible and no result is reported.}
  \label{tab:ev-results}
  \footnotesize 
  \setlength{\tabcolsep}{3.5pt}  
  \renewcommand{\arraystretch}{1.12}
  \begin{tabular*}{\textwidth}{@{\extracolsep{\fill}} c c c *{5}{c} *{5}{c} *{4}{c}@{}}
    \toprule
    \multicolumn{3}{c}{} &
    \multicolumn{5}{c}{\algCE} &
    \multicolumn{5}{c}{\algOP} &
    \multicolumn{4}{c}{\algEL} \\
    \cmidrule(lr){4-8}\cmidrule(lr){9-13}\cmidrule(lr){14-17}
    \textbf{$n$} & \textbf{$q$} & \textbf{$W_{\mathrm{tar}}$} &
    $\pi$ & $\lVert \underbar{E}\rVert$ & $\lVert \overbar{E}\rVert$ & $\mathcal{T}_1$ [s] & $\mathcal{T}_3$ [s] &
    $\pi$ & $\lVert \underbar{E}\rVert$ & $\lVert \overbar{E}\rVert$ & $\mathcal{T}_2$ [s] & $\mathcal{T}_3$ [s] &
    $\pi$ & $\lVert \underbar{E}\rVert$ & $\lVert \overbar{E}\rVert$ & $\mathcal{T}$ [s] \\
    \midrule
    6 & 1 & $W_1$       & 1 & 4.6817 & 4.6421 & 206.08 & 2.76 & 1 & 4.6817 & 4.6421 & 281.26 & 2.08 & 1 & 4.6817 & 4.6421 & 8.07 \\
    6 & 2 &  $W_2$      & 1 & 4.6968 & 4.7083 & 1091.07 & 3.08 & 1 & 4.6968 & 4.7083 & 1656.20 & 2.81 & 1 & 4.6968 & 4.7083 & 20.96 \\
    6 & 3 & $W_3$       & 1 & 4.7017 & 4.7241 & 2387.70 & 3.84 & 1 & 4.7017 & 4.7241 & 4059.09 & 3.18 & 1 & 4.7017 & 4.7241 & 51.68 \\
    6 & 1 &  $W_4$      & 15 & 3.4865 & 4.6175 & 328.41 & 3.08 & 16 & 2.8859 & 4.669 & 449.54 & 2.83 & 11 & 4.7259 & 4.4079 & 15.40 \\
    6 & 2 & $W_5$       & 15 & 3.5436 & 4.6541 & 1207.42 & 3.08 & 16 & 2.9546 & 4.7085 & 1922.10 & 3.19 & 1 & 14.8044 & 9.1324 & 103.11 \\
    6 & 3 &  $W_6$     & 15 & 3.5896 & 4.6852 & 2532.60 & 4.16 & 16 & 3.0453 & 4.7616 & 4478.60 & 4.14 & -- & -- & -- & -- \\
    6 & 1 &  $W_7$      & 2 & 2.8353 & 4.6752 & 234.73 & 1.87 & 4 & 2.3887 & 4.8574 & 316.89 & 1.12 & 3 & 2.9637 & 4.7976 & 7.42 \\
    6 & 2 & $W_8$       & 2 & 2.9483 & 4.8541 & 1018.42 & 2.12 & 4 & 2.4144 & 4.8831 & 1727.63 & 2.12 & 3 & 2.9845 & 4.8236 & 18.66 \\
    6 & 3 & $W_9$       & 3 & 2.9913 & 4.8291 & 2029.80 & 3.09 & 4 & 2.4247 & 4.8875 & 3485.20 & 3.09 & 3 & 2.9913 & 4.8291 & 69.37 \\
    \bottomrule
  \end{tabular*}
\end{table}

\smallskip
\noindent\textbf{Comparing Accuracies.}
Across all targets in the Table~\ref{tab:ev-results}, the PLF-based approaches \algCE and \algOP match or improve the accuracy of the baseline \algEL approach while selecting comparable policies. For $W_1$–$W_3$, the selected policy $\pi$ is identical across the three approaches, and the reported errors are the same. For $W_4$, $\lVert\underline{E}\rVert$ decreases under both \algCE and \algOP, whereas $\lVert\overline{E}\rVert$ increases slightly; however, the sum $\lVert\underline{E}\rVert+\lVert\overline{E}\rVert$ decreases, so the overall error band tightens. Over these rows, \algOP achieves a smaller total error than \algCE.
At $W_5$, the total error is larger than \algEL (i.e., the accuracy is reduced relative to the baseline), yet within the polyhedral class \algOP improves upon \algCE: $\lVert\underline{E}\rVert$ is smaller in \algOP, $\lVert\overline{E}\rVert$ is slightly larger in \algOP, and the sum is smaller in \algOP.

For $W_6$, the optimization in the baseline is infeasible, while both \algCE and \algOP return feasible certificates and policies; again, \algOP yields a smaller sum despite a slightly larger $\lVert\overline{E}\rVert$ and a smaller $\lVert\underline{E}\rVert$ than \algCE. 
Similarly for $W_7$ and $W_8$,
the total error is smaller in \algOP than \algCE, and smaller in \algCE than the baseline. %
Finally, at $W_9$, \algCE and \algEL select the same policy and attain the same errors, whereas \algOP selects a different policy with a
smaller total error. In summary, both polyhedral methods leverage geometric flexibility to tighten error bands relative to the baseline, and within the polyhedral family \algOP consistently attains the smallest total error, followed by \algCE.

\smallskip
\noindent
\textbf{Comparing Computational Times.}
The required time has two main elements: The PLF synthesis via Algorithms~\ref{alg learn cegis} or \ref{alg learn opt}, and the value computation in Algorithm~\ref{relax vi alg moimdp}. The computational times reported in Table~\ref{tab:ev-results} indicate that the value computation takes only a few seconds per instance, while the PLF synthesis stage dominates the overall runtime.
The \algEL baseline has the shortest execution time, followed by \algOP, with \algCE being the most time-consuming: \(\mathcal{T} < \mathcal{T}_2 + \mathcal{T}_3 < \mathcal{T}_1 + \mathcal{T}_3\). The higher computational time \(\mathcal{T}_1\) stems from the \algCE algorithm, which iteratively uses an SMT solver to synthesize and verify the satisfaction of the requirements. In contrast, \algOP (\(\mathcal{T}_2\)) alternates between solving MILPs, which tends to grow at a moderate rate. In summary, adopting the polyhedral ISoA enhances accuracy by leveraging geometric flexibility, but this improvement comes at the cost of longer synthesis times. Among the two polyhedral methods, \algOP achieves a better accuracy and requires less time than \algCE. 

\section{Conclusions} \label{sec: conclusion}
This paper presented a new approach to policy synthesis for multi-objective Markov decision processes using polyhedral Lyapunov functions. We reformulated the related value iteration algorithm as a switched affine system with interval uncertainties and applied control-theoretic theorems to synthesize policies that lead the system trajectories toward an invariant set that includes the target values of the objectives. The polyhedral function makes it suitable for the affine structure of the dynamic programming equations with improved accuracy in managing interval uncertainties.
Our method eliminated the need for costly Pareto-front computations or their approximations by augmenting different objectives in the switched affine system model. Numerical studies involving a recycling robot and an electric vehicle battery application demonstrated the effectiveness of our policy synthesis under uncertainty. 
Future work includes considering optimization problems that require solving constrained dynamic programming, and integrating abstraction methods with quantified error bounds for improving scalability by reducing the size of the model.

\bibliographystyle{cas-model2-names}

\bibliography{cas-refs}

\newpage
\appendix
\section{Proof of Theorem~\ref{thm cegis}}
\begin{pf}
The PLF $\V(E)$ being non-negative implies that
    \begin{align*}
        \forall E, \V(E)\ge \,0 \,\,\, \Leftrightarrow\,\,\, \forall E, \max_{c_i \in \Nu} \left({c_i}^\top E-d_{c_i}\right)\ge \,0 \,\,\, \Leftrightarrow\,\,\, \forall E,\, \exists c_i \in \Nu,\, {c_i}^\top E-d_{c_i} \ge \,0.
    \end{align*}
Hence, if the parameters satisfy condition $\psi_0$ as
\begin{equation}
    \psi_0: \forall E,\, \exists c_i \in \Nu,\, {c_i}^\top E-d_{c_i} \ge \,0, \label{eq pr psi 0}
\end{equation}
the PLF $\V(E)$ will be non-negative for all $E$. The negation of the condition $\psi_0$ is
\begin{equation}
    \neg \psi_0: \exists E,\, \forall c_i \in \Nu,\, {c_i}^\top E-d_{c_i} < 0. \label{eq pr neq psi 0}
\end{equation}
For the condition~(C1) in Definition~\ref{invariant set mdp def}, we have 
\begin{align*}
    0 \in \Omega\,\,\,\Leftrightarrow\,\,\, \V(0) \le 1 \,\,\,\Leftrightarrow\,\,\, \max_{c_i \in \Nu} (0 - d_{c_i}) \le 1 \,\,\,\Leftrightarrow\,\,\,\forall c_i \in \Nu, \, d_{c_i} \ge -1.
\end{align*}
Hence, condition~(C1) hold if the parameters satisfy condition $\psi_1$ as
\begin{equation}
    \psi_1: \forall c_i \in \Nu, \, d_{c_i} \ge -1. \label{eq pr psi 1}
\end{equation}
The negation of $\psi_1$ is
\begin{equation}
    \neg\psi_1: \exists c_i \in \Nu, \, d_{c_i} < -1. \label{eq pr neg psi 1}
\end{equation}
For the condition~(C2) in Definition~\ref{invariant set mdp def}, we have
\begin{align*}
&\V(E)>1 \ra \Delta \V(E) < 0\,\,\,
    \Leftrightarrow \,\,\, \max_{c_i \in \Nu} (c_i^\top E-d_{c_i}) > 1 \ra \\
    &\min_\pi \max_{c_j \in \Nu} (c_j^\top (A_\pi E + L_\pi)- d_{c_j}) - \max_{c_\ell \in \Nu} (c_\ell^\top E-d_{c_\ell})<0, 
\end{align*}
for all uncertain matrices $[A_\pi,L_\pi] \in \co ([A_\pi^\kappa,L_\pi^\kappa])_{\kappa \in \D}$. Equivalently, we have
\begin{align}
    &\forall [A_\pi,L_\pi] \in \co ([A_\pi^\kappa,L_\pi^\kappa])_{\kappa \in \D}, \,\, \max_{c_i \in \Nu} (c_i^\top E-d_{c_i}) \le 1 \,\, \vee \nonumber\\ &\min_\pi \max_{c_j \in \Nu} (c_j^\top (A_\pi E + L_\pi) < d_{c_j}) < \max_{c_\ell \in \Nu} (c_\ell^\top E-d_{c_\ell}). \label{eq pr c2} 
\end{align}
Based on~\eqref{eq pr c2} and the properties of convexity, the following condition $\psi_2$ satisfies the condition~(C2) in Definition~\ref{invariant set mdp def}:
\begin{align}
    &\psi_2: \forall E, \, \exists \pi, \, [\forall c_i \in \Nu, (c_i^\top E-d_{c_i}) \le 1] \,\vee\nonumber\\ 
    &[\forall c_j \in \Nu, \forall \kappa \in \D,\exists c_\ell \in \Nu, 
    (c_j^\top (A_\pi^\kappa E + L_\pi^\kappa) - d_{c_j} < c_\ell^\top E- d_{c_\ell})]. \label{eq pr psi 2}
\end{align}
The negation of the condition $\psi_2$ is
\begin{align}
    &\neg\psi_2: \exists E, \, \forall \pi, \, [\exists c_i \in \Nu, (c_i^\top E-d_{c_i}) > 1] \,\,\wedge \,\,
    [\exists c_j \in \Nu, \exists \kappa \in \D,\forall c_\ell \in \Nu, 
    (c_j^\top (A_\pi^\kappa E + L_\pi^\kappa) - d_{c_j} \ge c_\ell^\top E- d_{c_\ell})]. \label{eq pr neq psi 2}
\end{align}
Lastly, for the condition~(C3) in Definition~\ref{invariant set mdp def}, we have
\begin{align*}
 & \V(E) \le 1 \ra  \V(A_\pi E + L_\pi) \le 1 \,\,\,
    \Leftrightarrow \,\,\, \max_{c_i \in \Nu} (c_i^\top E -d_{c_i}) \leq 1 \\
    &\rightarrow  \min_\pi \max_{c_j \in \Nu} (c_j^\top (A_\pi E + L_\pi) -d_{c_j}) \leq 1,
\end{align*}
for all uncertain matrices $[A_\pi,L_\pi] \in \co ([A_\pi^\kappa,L_\pi^\kappa])_{\kappa \in \D}$.
This is equivalent to
\begin{align}
    &\forall [A_\pi,L_\pi] \in \co ([A_\pi^\kappa,L_\pi^\kappa])_{\kappa \in \D},\,\max_{c_i \in \Nu} (c_i^\top E -d_{c_i}) > 1 \,\, \vee \,\, \min_\pi \max_{c_j \in \Nu} (c_j^\top (A_\pi E + L_\pi) -d_{c_j}) \leq 1. \label{eq pr c3}
\end{align}
Utilizing~\eqref{eq pr c3} along with the convexity properties,
the following condition $\psi_3$ satisfies the condition~(C3) in Definition~\ref{invariant set mdp def}:
\begin{align}
    &\psi_3: \,\forall E, \exists \pi,\,\, [\exists c_i \in \Nu,\, (c_i^\top E -d_{c_i}) > 1] \,\, \vee \,\, [\forall \kappa \in \D,\,\forall c_j \in \Nu,\, c_j^\top (A_\pi^\kappa E + L_\pi^\kappa)-d_{c_j} \leq 1]. \label{eq pr psi3}
\end{align}
The negation of the condition $\psi_3$ is
\begin{align}
    \neg \psi_3: \,\exists E, \forall \pi,\,\, [\forall c_i \in \Nu,\, (c_i^\top E -d_{c_i}) \le 1] \,\, \wedge \,\, [\exists \kappa \in \D,\,\exists c_j \in \Nu,\, c_j^\top (A_\pi^\kappa E + L_\pi^\kappa)-d_{c_j} > 1]. \label{eq pr neq psi3}
\end{align}
Putting all the requirements of Definition~\ref{invariant set mdp def} together, we get   
\begin{align*}
    \Psi = \psi_0 \,\, \wedge \,\, \psi_1 \,\, \wedge \,\, \psi_2 \,\, \wedge \,\, \psi_3,
\end{align*}
which gives the expression in~\eqref{eq pr PSI}, with its negation being
\begin{equation*}
    \neg \Psi = \neg \psi_0 \, \vee \, \neg \psi_1 \, \vee \, \neg \psi_2 \, \vee \, \neg \psi_3,
\end{equation*}
which results in the expression~\eqref{eq pr neg Psi}, and concludes the proof.
$\hfill\blacksquare$ 
\end{pf}

\section{Proof of Theorem~\ref{thm Phi}}
\begin{pf}
    The PLF $\V(E)$ being non-negative implies that
    \begin{align*}
        \V(E) \ge 0 \,\, \Leftrightarrow\,\,\max_{c_i \in \Nu} {c_i}^\top E\ge 0 \,\,
        \Leftrightarrow\,\, \forall E,\, \exists c_i \in \Nu,\, {c_i}^\top E \ge 0.
    \end{align*}
Hence, if the parameters satisfy condition $\phi_0$ as
\begin{equation}
    \phi_0: \forall E,\, \exists c_i \in \Nu,\, {c_i}^\top E \ge 0, \label{eq pr phi 0}
\end{equation}
the PLF $\V(E)$ will be non-negative for all $E$. The negation of the condition $\phi_0$ is
\begin{equation}
    \neg \phi_0: \exists E,\, \forall c_i \in \Nu,\, {c_i}^\top E < 0. \label{eq pr neg phi 0}
\end{equation}
For the condition~(C1) in Definition~\ref{invariant set mdp def}, we have 
\begin{align*}
    0 \in \Omega \Ra \V(0) \le \rho &\Ra \max_{c_i \in \Nu} 0 \le \rho  \Ra 0 \le \rho,
\end{align*}
which is already satisfied by assuming $\rho \ge 0$.
For the condition~(C2) in Definition~\ref{invariant set mdp def}, we have
\begin{align*}
&\V(E)> \rho \ra \Delta \V(E) < 0 \,\,\, \Leftrightarrow \,\,\, \max_{c_i \in \Nu} (c_i^\top E) > \rho \ra 
     \min_\pi \max_{c_j \in \Nu} (c_j^\top (A_\pi E + L_\pi)) - \max_{c_\ell \in \Nu} (c_\ell^\top E) < 0,
\end{align*}
for all uncertain matrices $[A_\pi,L_\pi] \in \co ([A_\pi^\kappa,L_\pi^\kappa])_{\kappa \in \D}$.
Equivalently, we have
\begin{align}
    &\forall [A_\pi,L_\pi] \in \co ([A_\pi^\kappa,L_\pi^\kappa])_{\kappa \in \D}, \,\, \max_{c_i \in \Nu} (c_i^\top E) \le \rho \,\,\, \vee \,\,\, \min_\pi \max_{c_j \in \Nu} (c_j^\top (A_\pi E + L_\pi)) < \max_{c_\ell \in \Nu} (c_\ell^\top E). \label{eq pr c22} 
\end{align}
Using~\eqref{eq pr c22} and the convexity properties, 
the following condition $\phi_2$ satisfies the condition~(C2) in Definition~\ref{invariant set mdp def}:
\begin{align}
    &\phi_2: \forall E, \exists \pi, \, [\forall c_i \in \Nu, (c_i^\top E) \le \rho] \,\, \vee \,\,  
    [\forall c_j \in \Nu, \forall \kappa \in \D,
 \exists c_\ell \in \Nu, 
    (c_j^\top (A_\pi^\kappa E + L_\pi^\kappa) < c_\ell^\top E)]. \label{eq pr phi 2}
\end{align}
The negation of the condition $\phi_2$ is
\begin{align}
    &\neg\phi_2: \exists E, \forall \pi,\, [\exists c_i \in \Nu, (c_i^\top E) > \rho] \,\,\wedge 
    [\exists c_j \in \Nu, \exists \kappa \in \D, \forall c_\ell \in \Nu,
    (c_j^\top (A_\pi^\kappa E + L_\pi^\kappa) \ge c_\ell^\top E)]. \label{eq pr neg phi 2}
\end{align}
Lastly, For the condition~(C3) in Definition~\ref{invariant set mdp def}, we have
\begin{align*}
&\V(E) \le \rho \ra  \V(A_\pi E + L_\pi) \le \rho \,\,\, \Leftrightarrow \,\,\, \max_{c_i \in \Nu} c_i^\top E  \leq \rho \rightarrow \\ 
    &\min_\pi \max_{c_j \in \Nu} c_j^\top (A_\pi E + L_\pi) \leq \rho,
\end{align*}
for all uncertain matrices $[A_\pi,L_\pi] \in \co ([A_\pi^\kappa,L_\pi^\kappa])_{\kappa \in \D}$. 
Equivalently, we can express it as
\begin{align}
    &\forall [A_\pi,L_\pi] \in \co ([A_\pi^\kappa,L_\pi^\kappa])_{\kappa \in \D},\,\max_{c_i} c_i^\top E  > \rho \,\, \vee \,\, \min_\pi \max_{c_j \in \Nu} c_j^\top (A_\pi E + L_\pi) \leq \rho. \label{eq pr c33}
\end{align}
By applying~\eqref{eq pr c33} and the properties of convexity, condition \( \phi_3 \) below satisfies the condition~(C3) as specified in Definition~\ref{invariant set mdp def}:
\begin{align}
    \phi_3: \,\forall E,\exists \pi, \, [\exists c_i \in \Nu,\, (c_i^\top E) > \rho] \,\vee\, [\forall c_j \in \Nu,\,\,\forall \kappa \in \D,   c_j^\top (A_\pi^\kappa E + L_\pi^\kappa) \leq \rho]. \label{eq pr phi3}
\end{align}
The negation of the condition \(\phi_3\) is
\begin{align}
    \neg\phi_3: \,\exists E, \forall \pi, \, [\forall c_i \in \Nu,\, (c_i^\top E) \le \rho] \,\wedge\, [\exists c_j \in \Nu,\,\,\exists \kappa \in \D, c_j^\top (A_\pi^\kappa E + L_\pi^\kappa) > \rho]. \label{eq pr neg phi3}
\end{align}
Putting all the requirements of Definition~\ref{invariant set mdp def} together, we
get
\begin{align*}
    &\Phi = \phi_0 \,\, \wedge  \,\, \phi_2 \,\, \wedge \,\, \phi_3 = \forall E,\exists \pi, \{(\exists c_i \in \Nu,\, {c_i}^\top E \ge 0) \,\, \wedge \,\, ([\forall c_i \in \Nu, (c_i^\top E) \le \rho] \,\,\vee \\ 
    &[\forall c_j \in \Nu,\forall \kappa \in \D, \exists c_\ell \in \Nu, (c_j^\top (A_\pi^\kappa E + L_\pi^\kappa) < c_\ell^\top E)]) \,\, \wedge \,\, ([\exists c_i \in \Nu,\, (c_i^\top E) > \rho] \,\, \vee\nonumber\\
    &[\forall c_j \in \Nu,\,\,\forall \kappa \in \D,\,\, c_j^\top (A_\pi^\kappa E + L_\pi^\kappa) \leq \rho])\},
\end{align*}
Using the Boolean identity $(A \vee B)\wedge (\neg A \vee C) = (A \wedge C) \vee (\neg A \wedge B) $, we get
\begin{align*}
    &\Phi = \forall E, \exists \pi, (\exists c_i \in \Nu, {c_i}^\top E \ge 0)\,\wedge\,\{([\forall c_i \in \Nu, (c_i^\top E) \le \rho] \,\, \wedge \,\, [\forall c_j \in \Nu,\forall \kappa \in \D, c_j^\top (A_\pi^\kappa E + L_\pi^\kappa) \leq \rho]) \,\, \vee \\
    &([\exists c_i \in \Nu, (c_i^\top E) > \rho] \,\, \wedge \,\, [\forall c_j \in \Nu, \forall \kappa \in \D,\exists c_\ell \in \Nu, \,(c_j^\top (A_\pi^\kappa E + L_\pi^\kappa) < c_\ell^\top E)])\},
\end{align*}
which is the expression in~\eqref{Thm Phi cond}, with its negation being
\begin{equation*}
    \neg \Phi = \neg \phi_0 \, \vee \, \neg \phi_2 \, \vee \, \neg \phi_3,
\end{equation*}
which results in~\eqref{Thm neg Phi cond} using Boolean identities and concludes the proof.$\hfill\blacksquare$
\end{pf}
\end{document}